\documentclass[12pt]{article}
\usepackage{amsmath,amssymb}
\usepackage{graphicx}
\newcommand{\eqdef}{\stackrel{\text{def}}{=}}

\newcommand{\n}{\nonumber\\}
\newcommand{\bm}{\boldsymbol}
\newcommand{\ignore}[1]{}
\numberwithin{equation}{section}
\newcommand{\Romannumeral}[1]{\uppercase\expandafter{\romannumeral#1}}
\newcommand{\I}{\text{\Romannumeral{1}}}
\newcommand{\II}{\text{\Romannumeral{2}}}
\newcommand{\III}{\text{\Romannumeral{3}}}
\newcommand{\IV}{\text{\Romannumeral{4}}}
\newcommand{\V}{\text{\Romannumeral{5}}}
\newcommand{\VI}{\text{\Romannumeral{6}}}
\newtheorem{conj}{\bf Conjecture}%[section]
\allowdisplaybreaks[4]

\begin{document}

\baselineskip=20pt

%%%%%%%%%%%%%%%%%%%%%%%%%%%%%%%%%%%%%%%%%%%%%%%%%%%%%%%%%%%%
%                                                          %
%  Title page                                              %
%                                                          %
%%%%%%%%%%%%%%%%%%%%%%%%%%%%%%%%%%%%%%%%%%%%%%%%%%%%%%%%%%%%
\newfont{\elevenmib}{cmmib10 scaled\magstep1}
\newcommand{\preprint}{
    \begin{flushright}\normalsize \sf
     DPSU-19-1\\
%     {\tt arXiv:1907.12218[math-ph]}\\
%     July 2019
   \end{flushright}}
\newcommand{\Title}[1]{{\baselineskip=26pt
   \begin{center} \Large \bf #1 \\ \ \\ \end{center}}}
\newcommand{\Author}{\begin{center}
   \large \bf Satoru Odake \end{center}}
\newcommand{\Address}{\begin{center}
%     Department of Physics, Shinshu University,\\
     Faculty of Science, Shinshu University,\\
     Matsumoto 390-8621, Japan
   \end{center}}
\newcommand{\Accepted}[1]{\begin{center}
   {\large \sf #1}\\ \vspace{1mm}{\small \sf Accepted for Publication}
   \end{center}}

\preprint
\thispagestyle{empty}

\Title{Exactly Solvable Discrete Quantum Mechanical Systems and Multi-indexed
Orthogonal Polynomials of\\
the Continuous Hahn and Meixner-Pollaczek Types}

\Author

\Address
\vspace{1cm}

\begin{abstract}
We present new exactly solvable systems of the discrete quantum mechanics
with pure imaginary shifts, whose physical range of the coordinate is the
whole real line. These systems are shape invariant and their eigenfunctions
are described by the multi-indexed continuous Hahn and Meixner-Pollaczek
orthogonal polynomials.
The set of degrees of these multi-indexed polynomials are
$\{\ell_{\mathcal{D}},\ell_{\mathcal{D}}+1,\ell_{\mathcal{D}}+2,\ldots\}$,
where $\ell_{\mathcal{D}}$ is an even positive integer
($\mathcal{D}$ : a multi-index set), but
they form a complete set of orthogonal basis in the weighted Hilbert space.
\end{abstract}

%%%%%%%%%%%%%%%%%%%%%%%%%%%%%%%%%%%%%%%%%%%%%%%%%%%%%%%%%%%%%%%
%                                                             %
%  1. Introduction                                            %
%                                                             %
%%%%%%%%%%%%%%%%%%%%%%%%%%%%%%%%%%%%%%%%%%%%%%%%%%%%%%%%%%%%%%%
\section{Introduction}
\label{sec:intro}

Exactly solvable quantum mechanical systems in one dimension are closely
related to the orthogonal polynomials.
In ordinary quantum mechanics (oQM), whose Schr\"odinger equation is the
second order differential equation, the Hermite, Laguerre and Jacobi
polynomials appear in the harmonica oscillator, the radial oscillator and
the Darboux--P\"oschl--Teller potential, respectively.
In discrete quantum mechanics (dQM) \cite{os12,os13,os24},
whose Schr\"odinger equation is the second order difference equation,
the Askey-Wilson, $q$-Racah polynomials etc.\ appear.
Orthogonal polynomials satisfying second order differential or difference
equations are severely restricted by the Bochner's theorem and its
generalizations, and they are summarized as the Askey scheme of the (basic)
hypergeometric orthogonal polynomials \cite{ismail,kls}.
We have two types of dQM: dQM with pure imaginary shifts (idQM) and dQM with
real shifts (rdQM).
The coordinate of idQM is continuous and that of rdQM is discrete.

Recent developments in the theory of orthogonal polynomials and exactly
solvable quantum mechanical systems are based on the discovery of new types
of orthogonal polynomials: exceptional and multi-indexed polynomials
$\{P_{\mathcal{D},n}(\eta)|n\in\mathbb{Z}_{\geq 0}\}$
\cite{gkm08}--\cite{os35}.
These polynomials satisfy second order differential or difference
equations and form a complete set of orthogonal basis in an appropriate
Hilbert space in spite of missing degrees.
We distinguish the following two cases;
the set of missing degrees $\mathcal{I}=\mathbb{Z}_{\geq 0}\backslash
\{\text{deg}\,P_{\mathcal{D},n}(\eta)|n\in\mathbb{Z}_{\geq 0}\}$ is
case-(1): $\mathcal{I}=\{0,1,\ldots,\ell-1\}$, or
case-(2): $\mathcal{I}\neq\{0,1,\ldots,\ell-1\}$, where $\ell$ is a positive
integer. The situation of case-(1) is called stable in \cite{gkm11}.
Our approach to orthogonal polynomials is based on the quantum mechanical
formulation. We deform exactly solvable quantum mechanical systems by
multi-step Darboux transformations and obtain multi-indexed polynomials as
eigenfunctions of the deformed systems.
In the quantum mechanical formulation, the multi-indexed orthogonal
polynomials appear as polynomials in the sinusoidal coordinate $\eta(x)$
\cite{os7,os14}, $P_{\mathcal{D},n}(\eta(x))$, where $x$ is the
coordinate of the quantum system.

The range of the coordinate $x$ of oQM is a finite interval $(0,\frac12\pi)$
for the Darboux--P\"oschl--Teller potential (Jacobi polynomial),
the half real line $(0,\infty)$ for the radial oscillator (Laguerre polynomial)
and the whole real line $(-\infty,\infty)$ for the harmonic oscillator
(Hermite polynomial).
The counterparts of idQM to the Jacobi, Laguerre and Hermite polynomials of
oQM are Askey-Wilson, Wilson and continuous Hahn polynomials, respectively.
Their physical range of the coordinate $x$ is a finite interval $(0,\pi)$,
the half real line $(0,\infty)$ and the whole real line $(-\infty,\infty)$,
respectively \cite{kls}.
The situation of the multi-indexed polynomials for oQM and idQM constructed
so far is given in Table 1.
The case-(2) multi-indexed polynomials are obtained by taking the
eigenfunctions as seed solutions of the Darboux transformations,
and some eigenvalues are deleted from the original spectrum
\cite{crum}--\cite{gos}.
The Darboux transformations with the pseudo virtual state wavefunctions as
seed solutions also give the case-(2) multi-indexed polynomials, and
some eigenvalues are added to the original spectrum
\cite{q12b}--\cite{os30}.
%\cite{q12b,os29,os28,os30}.
The case-(1) multi-indexed polynomials are obtained by taking virtual state
wavefunctions as seed solutions of the Darboux transformations, and
the deformed systems are isospectral to the original systems
\cite{os25,os27}.
For rdQM systems, for example, see \cite{os22} for case-(2) and
\cite{os26,os35} for case-(1).

\begin{table}[htb]
\begin{center}
\renewcommand{\arraystretch}{1.2}
\begin{tabular}{|c|c|c|c|c|}
\hline
&(physical)&(typical)&
\multicolumn{2}{|c|}{multi-indexed polynomial}\\\cline{4-5}
&\ range of $x$\ &orthogonal polynomial
&\hspace{3mm}case-(1)\hspace*{3mm}&case-(2)\\\hline
&$(0,\frac12\pi)$&Jacobi&$\bigcirc$&$\bigcirc$\\\cline{2-5}
oQM&$(0,\infty)$&Laguerre&$\bigcirc$&$\bigcirc$\\\cline{2-5}
&$(-\infty,\infty)$&Hermite&{\Large$\times$}&$\bigcirc$\\\hline
&$(0,\pi)$&Askey-Wilson&$\bigcirc$&$\bigcirc$\\\cline{2-5}
idQM&$(0,\infty)$&Wilson&$\bigcirc$&$\bigcirc$\\\cline{2-5}
&$(-\infty,\infty)$&continuous Hahn&{\large ?}&$\bigcirc$\\\hline
\multicolumn{5}{r}{$\bigcirc$: possible and constructed,
{\Large$\times$}: impossible, {\large ?}: not yet studied}
\end{tabular}
\renewcommand{\arraystretch}{1.0}
\caption{multi-indexed polynomials in oQM and idQM}
\end{center}
\end{table}

The purpose of the present paper is to study the case-(1) multi-indexed
polynomials of idQM systems on the whole real line, namely, the case-(1)
multi-indexed polynomials of the continuous Hahn and Meixner-Pollaczek types,
which have not been studied except for the 1-indexed Meixner-Pollaczek
polynomial with special parameter \cite{os24}.
The case-(1) multi-indexed polynomials are obtained by the Darboux
transformations with the virtual state wavefunctions as seed solutions.
For oQM on the whole real line, there is no virtual state in the harmonic
oscillator and it is impossible to construct the case-(1) multi-indexed
Hermite polynomials.
For the (Askey-)Wilson cases studied in \cite{os27}, not only the final
deformed Hamiltonian but also the intermediate deformed Hamiltonians are
hermitian. For the continuous Hahn and Meixner-Pollaczek cases, however,
the intermediate deformed Hamiltonians may be singular.
This difference comes from the simple fact that an odd degree polynomial
with real coefficients has at least one zero on the whole real line.
We ignore the hermiticity of the intermediate deformed Hamiltonians and
require the hermiticity of the final deformed Hamiltonian only.

This paper is organized as follows.
In section \ref{sec:orgcHsys} the discrete quantum mechanics with pure
imaginary shifts is recapitulated and the data of the continuous Hahn
system is presented.
In section \ref{sec:newidQMcH} we deform the continuous Hahn idQM system
and obtain new exactly solvable idQM systems and the case-(1) multi-indexed
continuous Hahn polynomials.
In section \ref{sec:newidQMMP} we present the case-(1) multi-indexed
Meixner-Pollaczek polynomials and new exactly solvable idQM systems.
Section \ref{sec:summary} is for a summary and comments.
In Appendices \ref{app:prop} and \ref{app:MP:prop},
some properties of the multi-indexed continuous
Hahn and Meixner-Pollaczek polynomials are presented, respectively.

%%%%%%%%%%%%%%%%%%%%%%%%%%%%%%%%%%%%%%%%%%%%%%%%%%%%%%%%%%%%%%%
%                                                             %
%  2. Original Continuous Hahn System                         %
%                                                             %
%%%%%%%%%%%%%%%%%%%%%%%%%%%%%%%%%%%%%%%%%%%%%%%%%%%%%%%%%%%%%%%
\section{Original Continuous Hahn System}
\label{sec:orgcHsys}

After recapitulating the discrete quantum mechanics with pure imaginary shifts,
we present the data of the continuous Hahn system.

%%%%%%%%%%%%%%%%%%%%%%%%%%%%%%%%%%%%%%%%%%%%%
%                                           %
% 2.1 Discrete quantum mechanics            %
%     with pure imaginary shifts            %
%                                           %
%%%%%%%%%%%%%%%%%%%%%%%%%%%%%%%%%%%%%%%%%%%%%
\subsection{Discrete quantum mechanics with pure imaginary shifts}
\label{sec:idQM}

Let us recapitulate the discrete quantum mechanics with pure imaginary shifts
(idQM) \cite{os13,os24}.

The dynamical variables of idQM are the real coordinate $x$
%($x_1<x<x_2$)
($x_1\leq x\leq x_2$)
and the conjugate momentum $p=-i\partial_x$, which are governed by the
following factorized positive semi-definite Hamiltonian:
\begin{align}
  &\mathcal{H}\eqdef\sqrt{V(x)}\,e^{\gamma p}\sqrt{V^*(x)}
  +\!\sqrt{V^*(x)}\,e^{-\gamma p}\sqrt{V(x)}
  -V(x)-V^*(x)=\mathcal{A}^{\dagger}\mathcal{A},
  \label{H}\\
  &\mathcal{A}\eqdef i\bigl(e^{\frac{\gamma}{2}p}\sqrt{V^*(x)}
  -e^{-\frac{\gamma}{2}p}\sqrt{V(x)}\,\bigr),\quad
  \mathcal{A}^{\dagger}\eqdef-i\bigl(\sqrt{V(x)}\,e^{\frac{\gamma}{2}p}
  -\sqrt{V^*(x)}\,e^{-\frac{\gamma}{2}p}\bigr).
\end{align}
Here the potential function $V(x)$ is an analytic function of $x$ and
$\gamma$ is a real constant.
The $*$-operation on an analytic function $f(x)=\sum_na_nx^n$
($a_n\in\mathbb{C}$) is defined by $f^*(x)=\sum_na_n^*x^n$, in which
$a_n^*$ is the complex conjugation of $a_n$.
%Obviously $f^{**}(x)=f(x)$ and $f(x)^*=f^*(x^*)$.
%If a function satisfies $f^*=f$, then it takes real values on the real line.
Since the momentum operator appears in exponentiated forms,
the Schr\"{o}dinger equation
\begin{equation}
  \mathcal{H}\phi_n(x)=\mathcal{E}_n\phi_n(x)
  \ \ (n=0,1,2,\ldots),
  \label{Hphi=}
\end{equation}
is an analytic difference equation with pure imaginary shifts instead
of a differential equation.
Throughout this paper we consider those systems which have a
square-integrable groundstate together with an infinite number of
discrete energy levels:
$0=\mathcal{E}_0 <\mathcal{E}_1 < \mathcal{E}_2 < \cdots$.
The orthogonality relation reads
\begin{equation}
  (\phi_n,\phi_m)\eqdef
  \int_{x_1}^{x_2}\!\!dx\,\phi_n^*(x)\phi_m(x)=h_n\delta_{nm}
  \ \ (n,m=0,1,2,\ldots),\quad 0<h_n<\infty.
\end{equation}
The eigenfunctions $\phi_n(x)$ can be chosen `real', $\phi_n^*(x)=\phi_n(x)$,
and the groundstate wavefunction $\phi_0(x)$ is determined as the zero
mode of the operator $\mathcal{A}$, $\mathcal{A}\phi_0(x)=0$.
The norm of a function $f(x)$ is $|\!|f|\!|\eqdef(f,f)^{\frac12}$.

The Hamiltonian $\mathcal{H}$ should be hermitian.
{}From its form $\mathcal{H}=\mathcal{A}^{\dagger}\mathcal{A}$, it is formally
hermitian, $\mathcal{H}^{\dagger}=(\mathcal{A}^{\dagger}\mathcal{A})^{\dagger}
=(\mathcal{A})^{\dagger}(\mathcal{A}^{\dagger})^{\dagger}
=\mathcal{A}^{\dagger}\mathcal{A}=\mathcal{H}$.
However, the true hermiticity is defined in terms of the inner product,
$(f_1,\mathcal{H}f_2)=(\mathcal{H}f_1,f_2)$ \cite{os13,os14,os27}.
To show the hermiticity of $\mathcal{H}$, singularities of some functions
in the rectangular domain $D_{\gamma}$ are important.
Here $D_{\gamma}$ is defined by \cite{os27}
\begin{equation}
  D_{\gamma}\eqdef\bigl\{x\in\mathbb{C}\bigm|x_1\leq\text{Re}\,x\leq x_2,
  |\text{Im}\,x|\leq\tfrac12|\gamma|\bigr\}.
  \label{Dgamma}
\end{equation}

In the following, we assume that
the eigenfunctions $\phi_n(x)$ \eqref{Hphi=} have the following form:
\begin{equation}
  \phi_n(x)=\phi_0(x)\check{P}_n(x),\quad
  \check{P}_n(x)\eqdef P_n\bigl(\eta(x)\bigr)\quad
  (n=0,1,2,\ldots),
\end{equation}
where $\eta(x)$ is a sinusoidal coordinate \cite{os7,os14} and
$P_n(\eta)$ is a orthogonal polynomial of degree $n$ in $\eta$.
As a polynomial $P_n(\eta)$, we consider the Askey-Wilson, Wilson, continuous
Hahn polynomials etc., which are members of the Askey-scheme of hypergeometric
orthogonal polynomials \cite{kls}.
We call the idQM system by the name of the orthogonal polynomial:
Askey-Wilson system, Wilson system, continuous Hahn system etc.
These idQM systems have the property of shape invariance, which is a
sufficient condition for exact solvability.
Concrete idQM systems have a set of parameters
$\bm{\lambda}=(\lambda_1,\lambda_2,\ldots)$.
Various quantities depend on them and their dependence is expressed like,
$f=f(\bm{\lambda})$, $f(x)=f(x;\bm{\lambda})$.
(We sometimes omit writing $\bm{\lambda}$-dependence, when it does not
cause confusion.)

The shape invariant condition is the following \cite{os13,os14,os24}:
\begin{equation}
  \mathcal{A}(\bm{\lambda})\mathcal{A}(\bm{\lambda})^{\dagger}
  =\kappa\mathcal{A}(\bm{\lambda}+\bm{\delta})^{\dagger}
  \mathcal{A}(\bm{\lambda}+\bm{\delta})+\mathcal{E}_1(\bm{\lambda}),
  \label{shapeinv}
\end{equation}
where $\kappa$ is a real positive constant and $\bm{\delta}$ is the
shift of the parameters.
This condition combined with the Crum's theorem allows the wavefunction
$\phi_n(x)$ and energy eigenvalue $\mathcal{E}_n$ of the excited states to be
expressed in terms of the ground state wavefunction $\phi_0(x)$ and the first
excited state energy eigenvalue $\mathcal{E}_1$ with shifted parameters.
As a consequence of the shape invariance, we have
\begin{equation}
  \mathcal{A}(\bm{\lambda})\phi_n(x;\bm{\lambda})
  =f_n(\bm{\lambda})\phi_{n-1}(x;\bm{\lambda}+\bm{\delta}),\quad
  \mathcal{A}(\bm{\lambda})^{\dagger}\phi_{n-1}(x;\bm{\lambda}+\bm{\delta})
  =b_{n-1}(\bm{\lambda})\phi_n(x;\bm{\lambda}),
  \label{Aphi=,Adphi=}
\end{equation}
where $f_n(\bm{\lambda})$ and $b_{n-1}(\bm{\lambda})$ are some constants
satisfying
$f_n(\bm{\lambda})b_{n-1}(\bm{\lambda})=\mathcal{E}_n(\bm{\lambda})$.
These relations can be rewritten as
\begin{equation}
  \mathcal{F}(\bm{\lambda})\check{P}_n(x;\bm{\lambda})
  =f_n(\bm{\lambda})\check{P}_{n-1}(x;\bm{\lambda}+\bm{\delta}),\quad
  \mathcal{B}(\bm{\lambda})\check{P}_{n-1}(x;\bm{\lambda}+\bm{\delta})
  =b_{n-1}(\bm{\lambda})\check{P}_n(x;\bm{\lambda}).
  \label{FP=,BP=}
\end{equation}
Here the forward and backward shift operators $\mathcal{F}(\bm{\lambda})$ and
$\mathcal{B}(\bm{\lambda})$ are defined by
\begin{align}
  &\mathcal{F}(\bm{\lambda})\eqdef
  \phi_0(x;\bm{\lambda}+\bm{\delta})^{-1}\circ
  \mathcal{A}(\bm{\lambda})\circ\phi_0(x;\bm{\lambda})
  =i\varphi(x)^{-1}(e^{\frac{\gamma}{2}p}-e^{-\frac{\gamma}{2}p}),
  \label{Fdef}\\
  &\mathcal{B}(\bm{\lambda})\eqdef
  \phi_0(x;\bm{\lambda})^{-1}\circ
  \mathcal{A}(\bm{\lambda})^{\dagger}
  \circ\phi_0(x;\bm{\lambda}+\bm{\delta})
  =-i\bigl(V(x;\bm{\lambda})e^{\frac{\gamma}{2}p}
  -V^*(x;\bm{\lambda})e^{-\frac{\gamma}{2}p}\bigr)\varphi(x),
  \label{Bdef}
\end{align}
where $\varphi(x)$ is an auxiliary function
($\varphi(x)\propto\eta(x-i\frac{\gamma}{2})-\eta(x+i\frac{\gamma}{2})$).
%The second order difference operator $\widetilde{\mathcal{H}}(\bm{\lambda})$
The difference operator $\widetilde{\mathcal{H}}(\bm{\lambda})$
acting on the polynomial eigenfunctions is square root free:
\begin{align}
  &\widetilde{\mathcal{H}}(\bm{\lambda})\eqdef
  \phi_0(x;\bm{\lambda})^{-1}\circ\mathcal{H}(\bm{\lambda})
  \circ\phi_0(x;\bm{\lambda})
  =\mathcal{B}(\bm{\lambda})\mathcal{F}(\bm{\lambda})\n
  &\phantom{\widetilde{\mathcal{H}}_{\ell}(\bm{\lambda})}
  =V(x;\bm{\lambda})(e^{\gamma p}-1)
  +V^*(x;\bm{\lambda})(e^{-\gamma p}-1),\\
  &\widetilde{\mathcal{H}}(\bm{\lambda})\check{P}_n(x;\bm{\lambda})
  =\mathcal{E}_n(\bm{\lambda})\check{P}_n(x;\bm{\lambda}).
  \label{HtP=EP}
\end{align}

%%%%%%%%%%%%%%%%%%%%%%%%%%%%%%%%%%%%%%%%%%%%%%%
%                                             %
% 2.2 Continuous Hahn system                  %
%                                             %
%%%%%%%%%%%%%%%%%%%%%%%%%%%%%%%%%%%%%%%%%%%%%%%
\subsection{Continuous Hahn system}
\label{sec:cH}

Let us consider the continuous Hahn system.
The lower bound $x_1$, upper bound $x_2$ and the parameter $\gamma$ are
\begin{equation}
  x_1=-\infty,\quad x_2=\infty,\quad\gamma=1.
  \label{x1x2gamma}
\end{equation}
Namely, the physical range of the coordinate $x$ is a whole real line.
A set of parameters $\bm{\lambda}$ is
\begin{equation}
  \bm{\lambda}=(a_1,a_2),\quad a_i\in\mathbb{C},\quad\text{Re}\,a_i>0.
\end{equation}
Here are the fundamental data \cite{os13}:
\begin{align}
  &V(x;\bm{\lambda})=(a_1+ix)(a_2+ix)
  \ \ \bigl(\Rightarrow\,V^*(x;\bm{\lambda})=(a_1^*-ix)(a_2^*-ix)\bigr),
  \label{Vform}\\
  &\eta(x)=x,\ \ \varphi(x)=1,
  \ \ \mathcal{E}_n(\bm{\lambda})=n(n+b_1-1),
  \ \ b_1\eqdef a_1+a_2+a_1^*+a_2^*,\\
  &\phi_n(x;\bm{\lambda})
  =\phi_0(x;\bm{\lambda})\check{P}_n(x;\bm{\lambda}),
  \label{factphin}\\
  &\phi_0(x;\bm{\lambda})=
  \sqrt{\Gamma(a_1+ix)\Gamma(a_2+ix)\Gamma(a_1^*-ix)\Gamma(a_2^*-ix)},
  \label{phi0}\\
  &\check{P}_n(x;\bm{\lambda})=P_n\bigl(\eta(x);\bm{\lambda}\bigr)
  =p_n\bigl(\eta(x);a_1,a_2,a_1^*,a_2^*\bigr)
  \label{Pn}\\
  &\phantom{\check{P}_n(x;\bm{\lambda})}=
  i^n\frac{(a_1+a_1^*,a_1+a_2^*)_n}{n!}
  {}_3F_2\Bigl(\genfrac{}{}{0pt}{}{-n,\,n+b_1-1,\,a_1+ix}
  {a_1+a_1^*,\,a_1+a_2^*}\Bigm|1\Bigr)\\
  &\phantom{\check{P}_n(x;\bm{\lambda})}=
  c_n(\bm{\lambda})\eta(x)^n+(\text{lower order terms}),\quad
  c_n(\bm{\lambda})=\frac{(n+b_1-1)_n}{n!},
  \label{cn}\\
  &h_n(\bm{\lambda})=
  2\pi \frac{\prod_{j,k=1}^2\Gamma(n+a_j+a_k^*)}
  {n!\,(2n+b_1-1)\Gamma(n+b_1-1)},
  \label{hn}\\
  &\bm{\delta}=(\tfrac12,\tfrac12),\ \ \kappa=1,
  \ \ f_n(\bm{\lambda})=n+b_1-1,\ \ b_{n-1}(\bm{\lambda})=n.
\end{align}
(Although the notation $b_1$ conflicts with $b_{n-1}(\bm{\lambda})$,
we think this does not cause any confusion.)
Here $p_n(\eta;a_1,a_2,a_3,a_4)$ in \eqref{Pn} is the continuous Hahn
polynomial of degree $n$ in $\eta$ \cite{kls},
and the symbol $(a)_n$ is the shifted factorial.
Note that $\phi^*_0(x;\bm{\lambda})=\phi_0(x;\bm{\lambda})$
and $\check{P}^*_n(x;\bm{\lambda})=\check{P}_n(x;\bm{\lambda})$.
It is not necessary to distinguish $\check{P}_n$ and $P_n$ since $\eta(x)=x$,
but we will use both notations to compare with other cases in \cite{os27}.

%%%%%%%%%%%%%%%%%%%%%%%%%%%%%%%%%%%%%%%%%%%%%%%%%%%%%%%%%%%%%%%
%                                                             %
%  3. New Exactly Solvable idQM Systems and                   %
%     Multi-indexed Continuous Hahn Polynomials               %
%                                                             %
%%%%%%%%%%%%%%%%%%%%%%%%%%%%%%%%%%%%%%%%%%%%%%%%%%%%%%%%%%%%%%%
\section{New Exactly Solvable idQM Systems and Multi-indexed Continuous
Hahn Polynomials}
\label{sec:newidQMcH}

In this section we deform the continuous Hahn system by applying the multi-step
Darboux transformations with the virtual state wavefunctions as seed solutions.
The eigenfunctions of the deformed systems are described by the case-(1)
multi-indexed continuous Hahn polynomials.

%%%%%%%%%%%%%%%%%%%%%%%%%%%%%%%%%%%%%%%%%%%%%%%
%                                             %
% 3.1 Virtual state wavefunctions             %
%                                             %
%%%%%%%%%%%%%%%%%%%%%%%%%%%%%%%%%%%%%%%%%%%%%%%
\subsection{Virtual state wavefunctions}
\label{sec:vs}

Let us introduce two types of twist operations $\mathfrak{t}$ and constants
$\tilde{\bm{\delta}}$ :
\begin{align}
  \text{type $\I$}:&\ \ \ \mathfrak{t}^{\I}(\bm{\lambda})
  \eqdef(1-a_1^*,a_2),\quad
  \,\tilde{\bm{\delta}}^{\I}\eqdef(-\tfrac12,\tfrac12),\n
  \text{type $\II$}:&\ \ \mathfrak{t}^{\II}(\bm{\lambda})
  \eqdef(a_1,1-a_2^*),\quad
  \tilde{\bm{\delta}}^{\II}\eqdef(\tfrac12,-\tfrac12).
  \label{twist}
\end{align}
Each twist operation is an involution $\mathfrak{t}^2=\text{id}$, and
satisfies
$\mathfrak{t}(\bm{\lambda}+\beta\bm{\delta})=\mathfrak{t}(\bm{\lambda})
+\beta\tilde{\bm{\delta}}$ ($\beta\in\mathbb{R}$).
Their composition $\mathfrak{t}^{\III}(\bm{\lambda})\eqdef
(\mathfrak{t}^{\I}\circ\mathfrak{t}^{\II})(\bm{\lambda})
=(1-a_1^*,1-a_2^*)$ was used to construct the pseudo virtual
state wavefunctions \cite{os30}.
For each of $\mathfrak{t}^{\I}$ and $\mathfrak{t}^{\II}$, the potential
function $V(x;\bm{\lambda})$ satisfies
\begin{align}
  &V(x;\bm{\lambda})V^*(x-i\gamma;\bm{\lambda})
  =\alpha(\bm{\lambda})^2V\bigl(x;\mathfrak{t}(\bm{\lambda})\bigr)
  V^*\bigl(x-i\gamma;\mathfrak{t}(\bm{\lambda})\bigr),\n
  &V(x;\bm{\lambda})+V^*(x;\bm{\lambda})
  =\alpha(\bm{\lambda})\Bigl(V\bigl(x;\mathfrak{t}(\bm{\lambda})\bigr)
  +V^*\bigl(x;\mathfrak{t}(\bm{\lambda})\bigr)\Bigr)-\alpha'(\bm{\lambda}),
  \label{propV'}
\end{align}
where $\alpha(\bm{\lambda})$ and $\alpha'(\bm{\lambda})$ are
\begin{equation}
  \left\{\!\begin{array}{rcl}
  \alpha^{\I}(\bm{\lambda})&\!\!\!=\!\!\!&1\\[2pt]
  \alpha^{\II}(\bm{\lambda})&\!\!\!=\!\!\!&1
  \end{array}\right.,\quad
  \left\{\!\begin{array}{rcl}
  \alpha^{\prime\,\I}(\bm{\lambda})&\!\!\!=\!\!\!
  &-(a_1+a_1^*-1)(a_2+a_2^*)\\[2pt]
  \alpha^{\prime\,\II}(\bm{\lambda})&\!\!\!=\!\!\!
  &-(a_2+a_2^*-1)(a_1+a_1^*)
  \end{array}\right..
\end{equation}
In the following, we assume $\text{Re}\,a_i>\frac12$ ($i=1,2$),
which gives $\alpha'(\bm{\lambda})<0$.
The relations \eqref{propV'} imply a linear relation between two Hamiltonians
\cite{os27}:
\begin{equation}
  \mathcal{H}(\bm{\lambda})
  =\alpha(\bm{\lambda})\mathcal{H}\bigl(\mathfrak{t}(\bm{\lambda})\bigr)
  +\alpha'(\bm{\lambda}).
\end{equation}
%Since $\mathcal{H}(\bm{\lambda})$ is positive semi-definite,
%$\mathcal{H}\bigl(\mathfrak{t}(\bm{\lambda})\bigr)$ is obviously positive
%definite and it has no zero-mode.
Therefore $\phi_n(x;\mathfrak{t}(\bm{\lambda}))$ satisfies the Schr\"odinger
equation
$\mathcal{H}(\bm{\lambda})\phi_n(x;\mathfrak{t}(\bm{\lambda}))
=\tilde{\mathcal{E}}_n(\bm{\lambda})\phi_n(x;\mathfrak{t}(\bm{\lambda}))$
with $\tilde{\mathcal{E}}_n(\bm{\lambda})
=\alpha(\bm{\lambda})\mathcal{E}_n(\mathfrak{t}(\bm{\lambda}))
+\alpha'(\bm{\lambda})$.
Two types of virtual state wavefunctions
$\tilde{\phi}_{\text{v}}(x;\bm{\lambda})$
($\text{v}\in\mathcal{V}\subset\mathbb{Z}_{\geq 0}$) are defined by
\begin{align}
  \text{type $\I$}:&
  \ \ \tilde{\phi}^{\I}_{\text{v}}(x;\bm{\lambda})\eqdef
  \phi_{\text{v}}\bigl(x;\mathfrak{t}^{\I}(\bm{\lambda})\bigr)
  =\tilde{\phi}_0^{\I}(x;\bm{\lambda})
  \check{\xi}^{\I}_{\text{v}}(x;\bm{\lambda}),
  \ \ \tilde{\phi}^{\I}_0(x;\bm{\lambda})\eqdef
  \phi_0\bigl(x;\mathfrak{t}^{\I}(\bm{\lambda})\bigr),\n
  &\ \ \check{\xi}^{\I}_{\text{v}}(x;\bm{\lambda})\eqdef
  \xi^{\I}_{\text{v}}\bigl(\eta(x);\bm{\lambda}\bigr)\eqdef
  \check{P}_{\text{v}}\bigl(x;\mathfrak{t}^{\I}(\bm{\lambda})\bigr)
  =P_{\text{v}}\bigl(\eta(x);\mathfrak{t}^{\I}(\bm{\lambda})\bigr)
  \ \ (\text{v}\in\mathcal{V}^{\I}),\\
  \text{type $\II$}&
  \ \ \tilde{\phi}^{\II}_{\text{v}}(x;\bm{\lambda})\eqdef
  \phi_{\text{v}}\bigl(x;\mathfrak{t}^{\II}(\bm{\lambda})\bigr)
  =\tilde{\phi}_0^{\II}(x;\bm{\lambda})
  \check{\xi}^{\II}_{\text{v}}(x;\bm{\lambda}),
  \ \ \tilde{\phi}^{\II}_0(x;\bm{\lambda})\eqdef
  \phi_0\bigl(x;\mathfrak{t}^{\II}(\bm{\lambda})\bigr),\n
  &\ \ \check{\xi}^{\II}_{\text{v}}(x;\bm{\lambda})\eqdef
  \xi^{\II}_{\text{v}}\bigl(\eta(x);\bm{\lambda}\bigr)\eqdef
  \check{P}_{\text{v}}\bigl(x;\mathfrak{t}^{\II}(\bm{\lambda})\bigr)
  =P_{\text{v}}\bigl(\eta(x);\mathfrak{t}^{\II}(\bm{\lambda})\bigr)
  \ \ (\text{v}\in\mathcal{V}^{\II}).
  \label{xiv=}
\end{align}
The virtual state polynomials $\xi_{\text{v}}(\eta;\bm{\lambda})$
are polynomials of degree $\text{v}$ in $\eta$.
They are chosen `real,'
$\tilde{\phi}^*_0(x;\bm{\lambda})=\tilde{\phi}_0(x;\bm{\lambda})$,
$\check{\xi}^*_{\text{v}}(x;\bm{\lambda})
=\check{\xi}_{\text{v}}(x;\bm{\lambda})$
and the virtual energies $\tilde{\mathcal{E}}_{\text{v}}(\bm{\lambda})$ are
\begin{equation}
  \left\{\!\begin{array}{rcl}
  \tilde{\mathcal{E}}^{\I}_{\text{v}}(\bm{\lambda})
  &\!\!\!=\!\!\!&-(a_1+a_1^*-\text{v}-1)(a_2+a_2^*+\text{v})\\[4pt]
  \tilde{\mathcal{E}}^{\II}_{\text{v}}(\bm{\lambda})
  &\!\!\!=\!\!\!&-(a_2+a_2^*-\text{v}-1)(a_1+a_1^*+\text{v})
  \end{array}\right..
  \label{tEv}
\end{equation}
Note that $\alpha'(\bm{\lambda})=\tilde{\mathcal{E}}_0(\bm{\lambda})<0$ and
\begin{equation}
  \tilde{\mathcal{E}}^{\I}_{\text{v}}(\bm{\lambda})<0
  \ \Leftrightarrow\ a_1+a_1^*>\text{v}+1,\quad
  \tilde{\mathcal{E}}^{\II}_{\text{v}}(\bm{\lambda})<0
  \ \Leftrightarrow\ a_2+a_2^*>\text{v}+1,
  \label{tEv<0}
\end{equation}
for $\text{v}\geq 0$.
We choose $\mathcal{V}^{\I}$ and $\mathcal{V}^{\II}$ as
\begin{equation}
  \mathcal{V}^{\I}=\bigl\{0,1,2,\ldots,[a_1+a_1^*-1]'\bigr\},\quad
  \mathcal{V}^{\II}=\bigl\{0,1,2,\ldots,[a_2+a_2^*-1]'\bigr\},
  \label{vrange}
\end{equation}
where $[x]'$ denotes the greatest integer not equal or exceeding $x$.
Although we have included 0 in $\mathcal{V}$, the Darboux transformations with
the label 0 virtual state do not give essentially new systems, see the end of
\S\,\ref{sec:shapeinv}.

%%%%%%%%%%%%%%%%%%%%%%%%%%%%%%%%%%%%%%%%%%%%%%%
%                                             %
% 3.2 New exactly solvable systems            %
%                                             %
%%%%%%%%%%%%%%%%%%%%%%%%%%%%%%%%%%%%%%%%%%%%%%%
\subsection{New exactly solvable systems}
\label{sec:newsys}

By applying multi-step Darboux transformations to the continuous Hahn system
in \S\,\ref{sec:cH}, we can deform it and obtain new exactly solvable idQM
systems. The virtual state wavefunctions in \S\,\ref{sec:vs} are used as seed
solutions, and new systems are isospectral to the original one.

The deformed systems are labeled by
$\mathcal{D}=\{d_1,\ldots,d_M\}=\{d^{\I}_1,\ldots,d^{\I}_{M_{\I}},
d^{\II}_1,\ldots,d^{\II}_{M_{\II}}\}$ ($M=M_{\I}+M_{\II}$,
$d^{\I}_j\in\mathcal{V}^{\I}$ : mutually distinct,
$d^{\II}_j\in\mathcal{V}^{\II}$ : mutually distinct),
which are the degrees and types of the virtual state wavefunctions used in
$M$-step Darboux transformations.
The Hamiltonian is deformed as
$\mathcal{H}\to\mathcal{H}_{d_1}\to\mathcal{H}_{d_1d_2}\to\cdots\to
\mathcal{H}_{d_1\ldots d_s}\to\cdots\to
\mathcal{H}_{d_1\ldots d_M}=\mathcal{H}_{\mathcal{D}}$
by $M$-step Darboux transformations.
Exactly speaking, $\mathcal{D}$ is an ordered set.
Various quantities of the deformed systems are denoted as
%$\mathcal{H}_{d_1\ldots d_M}$, $\phi_{d_1\ldots d_M\,n}$,
%$\mathcal{A}_{d_1\ldots d_M}$, etc. or simply
$\mathcal{H}_{\mathcal{D}}$, $\phi_{\mathcal{D}\,n}$,
$\mathcal{A}_{\mathcal{D}}$, etc.
The general formula is as follows \cite{os27}:
\begin{align}
  &\mathcal{H}_{\mathcal{D}}\phi_{\mathcal{D}\,n}(x)
  =\mathcal{E}_n\phi_{\mathcal{D}\,n}(x)\ \ (n=0,1,2,\ldots),\\
  &\mathcal{H}_{\mathcal{D}}
  =\mathcal{A}_{\mathcal{D}}^{\dagger}\mathcal{A}_{\mathcal{D}},
  \label{HD}\\
  &\mathcal{A}_{\mathcal{D}}=
  i\bigl(e^{\frac{\gamma}{2}p}\sqrt{V_{\mathcal{D}}^*(x)}
  -e^{-\frac{\gamma}{2}p}\sqrt{V_{\mathcal{D}}(x)}\,\bigr),\quad
  \mathcal{A}_{\mathcal{D}}^{\dagger}=
  -i\bigl(\sqrt{V_{\mathcal{D}}(x)}\,e^{\frac{\gamma}{2}p}
  -\sqrt{V_{\mathcal{D}}^*(x)}\,e^{-\frac{\gamma}{2}p}\bigr),
  \label{AD}\\
  &V_{\mathcal{D}}(x)=
  \sqrt{V(x-i\tfrac{M}{2}\gamma)V^*(x-i\tfrac{M+2}{2}\gamma)}\n
  &\phantom{V_{\mathcal{D}}(x)=}\times
  \frac{\text{W}_{\gamma}[\tilde{\phi}_{d_1},\ldots,\tilde{\phi}_{d_M}]
  (x+i\frac{\gamma}{2})}
  {\text{W}_{\gamma}[\tilde{\phi}_{d_1},\ldots,\tilde{\phi}_{d_M}]
  (x-i\frac{\gamma}{2})}\,
  \frac{\text{W}_{\gamma}[\tilde{\phi}_{d_1},\ldots,\tilde{\phi}_{d_M},
  \phi_0](x-i\gamma)}
  {\text{W}_{\gamma}[\tilde{\phi}_{d_1},\ldots,\tilde{\phi}_{d_M},\phi_0](x)},
  \label{VD}\\
  &\phi_{\mathcal{D}\,n}(x)=
  \text{W}_{\gamma}[\tilde{\phi}_{d_1},\ldots,\tilde{\phi}_{d_M},\phi_n](x)\n
  &\phantom{\phi_{\mathcal{D}\,n}(x)=}
  \times\left(
  \frac{\sqrt{\prod_{j=0}^{M-1}V(x+i(\frac{M}{2}-j)\gamma)
  V^*(x-i(\frac{M}{2}-j)\gamma)}}
  {\text{W}_{\gamma}[\tilde{\phi}_{d_1},\ldots,\tilde{\phi}_{d_M}]
  (x-i\frac{\gamma}{2})
  \text{W}_{\gamma}[\tilde{\phi}_{d_1},\ldots,\tilde{\phi}_{d_M}]
  (x+i\frac{\gamma}{2})}\right)^{\frac12},
  \label{phiDn}
\end{align}
where $\text{W}_{\gamma}[f_1,\ldots,f_n]$ is the Casorati determinant of
a set of $n$ functions $\{f_j(x)\}$,
\begin{equation}
  \text{W}_{\gamma}[f_1,\ldots,f_n](x)
  \eqdef i^{\frac12n(n-1)}
  \det\Bigl(f_k\bigl(x^{(n)}_j\bigr)\Bigr)_{1\leq j,k\leq n},\quad
  x_j^{(n)}\eqdef x+i(\tfrac{n+1}{2}-j)\gamma,
  \label{Wdef}
\end{equation}
(for $n=0$, we set $\text{W}_{\gamma}[\cdot](x)=1$).
These properties of the Darboux transformation are proved algebraically,
% in an algebraic way
and their analytical properties are not considered. Therefore, the deformed
Hamiltonian $\mathcal{H}_{\mathcal{D}}$ may be singular.
In order to obtain well-defined deformed systems, we have to check the
regularity and hermiticity of $\mathcal{H}_{\mathcal{D}}$.
It is also necessary to check the square integrability of
$\phi_{\mathcal{D}\,n}(x)$.

To obtain the concrete forms of $V_{\mathcal{D}}(x)$ and
$\phi_{\mathcal{D}\,n}(x)$, we have to evaluate the Casoratians.
Let us define the following functions:
\begin{align}
  &\nu^{\I}(x;\bm{\lambda})\eqdef
  \frac{\phi_0(x;\bm{\lambda})}{\tilde{\phi}^{\I}_0(x;\bm{\lambda})},
  \ \ r_j^{\I}(x^{(M)}_j;\bm{\lambda},M)\eqdef
  \frac{\nu^{\I}(x^{(M)}_j;\bm{\lambda})}
  {\nu^{\I}\bigl(x;\bm{\lambda}+(M-1)\tilde{\bm{\delta}}^{\I}\bigr)}
  \ \ (j=1,2,\ldots,M),\n
  &\nu^{\II}(x;\bm{\lambda})\eqdef
  \frac{\phi_0(x;\bm{\lambda})}{\tilde{\phi}^{\II}_0(x;\bm{\lambda})},
  \ \ r_j^{\II}(x^{(M)}_j;\bm{\lambda},M)\eqdef
  \frac{\nu^{\II}(x^{(M)}_j;\bm{\lambda})}
  {\nu^{\II}\bigl(x;\bm{\lambda}+(M-1)\tilde{\bm{\delta}}^{\II}\bigr)}
  \ \ (j=1,2,\ldots,M),
\end{align}
whose explicit forms are
\begin{align}
  r^{\I}_j(x^{(M)}_j;\bm{\lambda},M)
  &=(-1)^{j-1}i^{1-M}(a_1-\tfrac{M-1}{2}+ix)_{j-1}
  (a_1^*-\tfrac{M-1}{2}-ix)_{M-j},\n
  r^{\II}_j(x^{(M)}_j;\bm{\lambda},M)
  &=(-1)^{j-1}i^{1-M}(a_2-\tfrac{M-1}{2}+ix)_{j-1}
  (a_2^*-\tfrac{M-1}{2}-ix)_{M-j}.
\end{align}
Furthermore, let us define $\check{\Xi}_{\mathcal{D}}(x;\bm{\lambda})$ and
$\check{P}_{\mathcal{D},n}(x;\bm{\lambda})$ as follows:
\begin{align}
  &i^{\frac12M(M-1)}\left|
  \begin{array}{llllll}
  \vec{X}^{(M)}_{d^{\I}_1}&\cdots&\vec{X}^{(M)}_{d^{\I}_{M_{\I}}}&
  \vec{Y}^{(M)}_{d^{\II}_1}&\cdots&\vec{Y}^{(M)}_{d^{\II}_{M_{\II}}}\\
  \end{array}\right|
  =\varphi_M(x)\check{\Xi}_{\mathcal{D}}(x;\bm{\lambda})\times A,
  \label{cXiDdef}\\
  &i^{\frac12M(M+1)}\left|
  \begin{array}{lllllll}
  \vec{X}^{(M+1)}_{d^{\I}_1}&\cdots&\vec{X}^{(M+1)}_{d^{\I}_{M_{\I}}}&
  \vec{Y}^{(M+1)}_{d^{\II}_1}&\cdots&\vec{Y}^{(M+1)}_{d^{\II}_{M_{\II}}}&
  \vec{Z}^{(M+1)}_n\\
  \end{array}\right|\n
  &=\varphi_{M+1}(x)\check{P}_{\mathcal{D},n}(x;\bm{\lambda})\times B,
  \label{cPDndef}
\end{align}
where $A$ and $B$ are
\begin{align}
  &A=\prod_{j=1}^{M_{\I}-1}
  (a_2-\tfrac{M-1}{2}+ix,a_2^*-\tfrac{M-1}{2}-ix)_j
  \cdot\prod_{j=1}^{M_{\II}-1}
  (a_1-\tfrac{M-1}{2}+ix,a_1^*-\tfrac{M-1}{2}-ix)_j,
  \label{cXiDA}\\
  &B=\prod_{j=1}^{M_{\I}}
  (a_2-\tfrac{M}{2}+ix,a_2^*-\tfrac{M}{2}-ix)_j
  \cdot\prod_{j=1}^{M_{\II}}
  (a_1-\tfrac{M}{2}+ix,a_1^*-\tfrac{M}{2}-ix)_j,
  \label{cPDnB}
\end{align}
and $\vec{X}^{(M)}_{\text{v}}$, $\vec{Y}^{(M)}_{\text{v}}$ and
$\vec{Z}^{(M)}_{\text{v}}$ are
\begin{align}
  &\bigl(\vec{X}^{(M)}_{\text{v}}\bigr)_j
  =r^{\II}_j(x^{(M)}_j;\bm{\lambda},M)
  \check{\xi}^{\I}_{\text{v}}(x^{(M)}_j;\bm{\lambda}),\qquad
  (j=1,2,\ldots,M),\n
  &\bigl(\vec{Y}^{(M)}_{\text{v}}\bigr)_j
  =r^{\I}_j(x^{(M)}_j;\bm{\lambda},M)
  \check{\xi}^{\II}_{\text{v}}(x^{(M)}_j;\bm{\lambda}),\n
  &\bigl(\vec{Z}^{(M)}_n\bigr)_j
  =r^{\II}_j(x^{(M)}_j;\bm{\lambda},M)r^{\I}_j(x^{(M)}_j;\bm{\lambda},M)
  \check{P}_n(x^{(M)}_j;\bm{\lambda}).
\end{align}
The auxiliary function $\varphi_M(x)$ introduced in \cite{gos} is
$\varphi_M(x)=1$ in the present case.
These $\check{\Xi}_{\mathcal{D}}(x;\bm{\lambda})$ and
$\check{P}_{\mathcal{D},n}(x;\bm{\lambda})$ are `real',
$\check{\Xi}^*_{\mathcal{D}}(x;\bm{\lambda})
=\check{\Xi}_{\mathcal{D}}(x;\bm{\lambda})$ and
$\check{P}^*_{\mathcal{D},n}(x;\bm{\lambda})
=\check{P}_{\mathcal{D},n}(x;\bm{\lambda})$.
They are polynomials in the sinusoidal coordinate $\eta(x)$:
\begin{equation}
  \check{\Xi}_{\mathcal{D}}(x;\bm{\lambda})\eqdef
  \Xi_{\mathcal{D}}\bigl(\eta(x);\bm{\lambda}\bigr),\quad
  \check{P}_{\mathcal{D},n}(x;\bm{\lambda})\eqdef
  P_{\mathcal{D},n}\bigl(\eta(x);\bm{\lambda}\bigr).
  \label{XiP_poly}
\end{equation}
We call $\Xi_{\mathcal{D}}(\eta;\bm{\lambda})$ the denominator polynomial and
$P_{\mathcal{D},n}(\eta;\bm{\lambda})$ the multi-indexed polynomial.
Their degrees are $\ell_{\mathcal{D}}$ and $\ell_{\mathcal{D}}+n$,
respectively (we assume $c_{\mathcal{D}}^{\Xi}(\bm{\lambda})\neq 0$ and
$c_{\mathcal{D},n}^{P}(\bm{\lambda})\neq 0$, see \eqref{cXiD}--\eqref{cPDn}).
Here $\ell_{\mathcal{D}}$ is
\begin{equation}
  \ell_{\mathcal{D}}\eqdef\sum_{j=1}^Md_j
  -\tfrac12M(M-1)+2M_{\I}M_{\II}.
  \label{lD}
\end{equation}
Then, the Casoratians 
$\text{W}_{\gamma}[\tilde{\phi}_{d_1},\ldots,\tilde{\phi}_{d_M}](x)$ and
$\text{W}_{\gamma}[\tilde{\phi}_{d_1},\ldots,\tilde{\phi}_{d_M},\phi_n](x)$
are expressed as
\begin{align}
  &\quad\text{W}_{\gamma}[\tilde{\phi}_{d_1},\ldots,\tilde{\phi}_{d_M}]
  (x;\bm{\lambda})\n
  &=\prod_{j=1}^M\phi_0\bigl(x^{(M)}_j;\bm{\lambda}\bigr)\cdot
  \text{W}_{\gamma}\bigl[\frac{1}{\nu^{\I}}\check{\xi}^{\I}_{d^{\I}_1},\ldots,
  \frac{1}{\nu^{\I}}\check{\xi}^{\I}_{d^{\I}_{M_{\I}}},
  \frac{1}{\nu^{\II}}\check{\xi}^{\II}_{d^{\II}_1},\ldots,
  \frac{1}{\nu^{\II}}\check{\xi}^{\II}_{d^{\II}_{M_{\II}}}
  \bigr](x;\bm{\lambda})\n
  &=\prod_{j=1}^M\phi_0\bigl(x^{(M)}_j;\bm{\lambda}\bigr)\cdot
  \nu^{\I}\bigl(x;\bm{\lambda}+(M-1)\tilde{\bm{\delta}}^{\I}\bigr)^{-M_{\I}}
  \nu^{\II}\bigl(x;\bm{\lambda}
  +(M-1)\tilde{\bm{\delta}}^{\II}\bigr)^{-M_{\II}}\n
  &\quad\times\prod_{j=1}^{M+1}
  r^{\I}_j\bigl(x^{(M)}_j;\bm{\lambda},M\bigr)^{-1}
  r^{\II}_j\bigl(x^{(M)}_j;\bm{\lambda},M\bigr)^{-1}
  \times\varphi_M(x)\check{\Xi}_{\mathcal{D}}(x;\bm{\lambda})A,\\
  &\quad\text{W}_{\gamma}[\tilde{\phi}_{d_1},\ldots,\tilde{\phi}_{d_M},\phi_n]
  (x;\bm{\lambda})\n
  &=\prod_{j=1}^{M+1}\phi_0\bigl(x^{(M+1)}_j;\bm{\lambda}\bigr)\cdot
  \text{W}_{\gamma}\bigl[\frac{1}{\nu^{\I}}\check{\xi}^{\I}_{d^{\I}_1},\ldots,
  \frac{1}{\nu^{\I}}\check{\xi}^{\I}_{d^{\I}_{M_{\I}}},
  \frac{1}{\nu^{\II}}\check{\xi}^{\II}_{d^{\II}_1},\ldots,
  \frac{1}{\nu^{\II}}\check{\xi}^{\II}_{d^{\II}_{M_{\II}}},
  \check{P}_n\bigr](x;\bm{\lambda})\n
  &=\prod_{j=1}^{M+1}\phi_0\bigl(x^{(M+1)}_j;\bm{\lambda}\bigr)\cdot
  \nu^{\I}\bigl(x;\bm{\lambda}+M\tilde{\bm{\delta}}^{\I}\bigr)^{-M_{\I}}
  \nu^{\II}\bigl(x;\bm{\lambda}+M\tilde{\bm{\delta}}^{\II}\bigr)^{-M_{\II}}\n
  &\quad\times\prod_{j=1}^{M+1}
  r^{\I}_j\bigl(x^{(M+1)}_j;\bm{\lambda},M+1\bigr)^{-1}
  r^{\II}_j\bigl(x^{(M+1)}_j;\bm{\lambda},M+1\bigr)^{-1}
  \times\varphi_{M+1}(x)\check{P}_{\mathcal{D},n}(x;\bm{\lambda})B,
\end{align}
where $A$ and $B$ are given in \eqref{cXiDA} and \eqref{cPDnB} respectively.
After some calculation, the eigenfunction \eqref{phiDn} is rewritten as
\begin{align}
  \phi_{\mathcal{D}\,n}(x;\bm{\lambda})&=
  \psi_{\mathcal{D}}(x;\bm{\lambda})
  \check{P}_{\mathcal{D},n}(x;\bm{\lambda}),
  \label{phiDn2}\\
  \psi_{\mathcal{D}}(x;\bm{\lambda})&\eqdef
  \frac{\phi_0(x;\bm{\lambda}^{[M_{\I},M_{\II}]})}
  {\sqrt{\check{\Xi}_{\mathcal{D}}(x-i\frac{\gamma}{2};\bm{\lambda})
  \check{\Xi}_{\mathcal{D}}(x+i\frac{\gamma}{2};\bm{\lambda})}},\quad
  \bm{\lambda}^{[M_{\I},M_{\II}]}\eqdef
  \bm{\lambda}+M_{\I}\tilde{\bm{\delta}}^{\I}+M_{\II}\tilde{\bm{\delta}}^{\II}.
  \label{psiD}
\end{align}
The ground state wavefunction $\phi_{\mathcal{D}\,0}$ is annihilated by
$\mathcal{A}_{\mathcal{D}}$,
$\mathcal{A}_{\mathcal{D}}(\bm{\lambda})
\phi_{\mathcal{D}\,0}(x;\bm{\lambda})=0$.
The lowest degree multi-indexed orthogonal polynomial
$\check{P}_{\mathcal{D},0}(x;\bm{\lambda})$ is proportional to
$\check{\Xi}_{\mathcal{D}}(x;\bm{\lambda}+\bm{\delta})$, see \eqref{PD0=A.XiD}.
The potential function $V_{\mathcal{D}}(x)$ \eqref{VD} is expressed neatly
in terms of the denominator polynomial:
\begin{equation}
  V_{\mathcal{D}}(x;\bm{\lambda})
  =V(x;\bm{\lambda}^{[M_{\I},M_{\II}]})\,
  \frac{\check{\Xi}_{\mathcal{D}}(x+i\frac{\gamma}{2};\bm{\lambda})}
  {\check{\Xi}_{\mathcal{D}}(x-i\frac{\gamma}{2};\bm{\lambda})}
  \frac{\check{\Xi}_{\mathcal{D}}(x-i\gamma;\bm{\lambda}+\bm{\delta})}
  {\check{\Xi}_{\mathcal{D}}(x;\bm{\lambda}+\bm{\delta})}.
  \label{VD2}
\end{equation}
Since the deformed Hamiltonian $\mathcal{H}_{\mathcal{D}}(\bm{\lambda})$
is expressed in terms of the potential function
$V_{\mathcal{D}}(x;\bm{\lambda})$, $\mathcal{H}_{\mathcal{D}}(\bm{\lambda})$
is determined by the denominator polynomial
$\check{\Xi}_{\mathcal{D}}(x;\bm{\lambda})$, whose normalization is irrelevant.
Under the permutation of $d_j$'s, the deformed Hamiltonian
$\mathcal{H}_{\mathcal D}$ is invariant, but the denominator polynomial
$\check{\Xi}_{\mathcal{D}}(x)$ and the multi-indexed polynomials
$\check{P}_{\mathcal{D},n}(x)$ may change their signs.

As mentioned before, we have to check the regularity and hermiticity of
$\mathcal{H}_{\mathcal{D}}(\bm{\lambda})$.
Let us consider the function $g(x)$,
\begin{align}
  g(x)&\eqdef V(x+i\tfrac{\gamma}{2};\bm{\lambda}^{[M_{\I},M_{\II}]})
  \phi_0(x+i\tfrac{\gamma}{2};\bm{\lambda}^{[M_{\I},M_{\II}]})^2\n
  &=\Gamma(a_1-\tfrac12M'+\tfrac12+ix)\Gamma(a_2+\tfrac12M'+\tfrac12+ix)\n
  &\quad\times
  \Gamma(a_1^*-\tfrac12M'+\tfrac12-ix)\Gamma(a_2^*+\tfrac12M'+\tfrac12-ix),
\end{align}
where $M'=M_{\I}-M_{\II}$.
Asymptotic behavior of $g(x)$ at $x\sim\pm\infty$ is
%$g(x)\sim 4\pi^2e^{\mp\pi\text{Im}\,(a_1+a_2)}
%|x|^{2\text{Re}\,(a_1+a_2)}e^{-2\pi|x|}$
$g(x)\sim 4\pi^2e^{\mp\pi\text{Im}\,(a_1+a_2)}$
$|x|^{2\text{Re}\,(a_1+a_2)}e^{-2\pi|x|}$
(for $x\in\mathbb{R}$),
where we have used the asymptotic formula of the gamma function
$|\Gamma(x+iy)|^2\sim 2\pi|y|^{2x-1}e^{-\pi|y|}$
($x,y\in\mathbb{R}$, $x$:\,fixed, $y\sim\pm\infty$).
The necessary and sufficient condition for $g(x)$ to have no poles in the
rectangular domain $D_{\gamma}$ is
$\text{Re}\,a_1-\frac12M'>0$ and $\text{Re}\,a_2+\frac12M'>0$.
This condition is automatically satisfied, because \eqref{vrange} implies
$M_{\I}-1\leq\max_j\{d^{\I}_j\}<2\text{Re}\,a_1-1$ and
$M_{\II}-1\leq\max_j\{d^{\II}_j\}<2\text{Re}\,a_2-1$, for $M_{\I},M_{\II}>0$.
(For $M_{\I}=0$ or $M_{\II}=0$, it is trivial.)
By the same argument as ref.\cite{os27} (\S\,2.2 and \S\,3.4),
the deformed Hamiltonian $\mathcal{H}_{\mathcal{D}}(\bm{\lambda})$ is
well-defined and hermitian, if the following condition is satisfied:
\begin{equation}
  \text{The denominator polynomial $\check{\Xi}_{\mathcal{D}}(x;\bm{\lambda})$
  has no zero in $D_{\gamma}$ \eqref{Dgamma}.}
  \label{nozero}
\end{equation}
This is a sufficient condition for the hermiticity.
For the Wilson and Askey-Wilson cases \cite{os27}, not only the final $M$-step
Hamiltonian $\mathcal{H}_{\mathcal{D}}$ but also the intermediate $s$-step
Hamiltonians $\mathcal{H}_{d_1\ldots d_s}$ are well-defined and hermitian.
In the present case, however, the intermediate $s$-step Hamiltonians
$\mathcal{H}_{d_1\ldots d_s}$ may be singular.
This situation is similar to the $M$-step Darboux transformations with
the eigenfunctions as seed solutions \cite{gos}, in which the intermediate
$s$-step Hamiltonians may be singular but the final $M$-step Hamiltonian
is well-defined and hermitian if the Krein-Adler condition is satisfied:
$\prod_{j=1}^M(n-d_j)\geq0$ ($\forall n\in\mathbb{Z}_{\geq 0}$).
To satisfy the condition \eqref{nozero}, the degree of
$\Xi_{\mathcal{D}}(\eta;\bm{\lambda})$, $\ell_{\mathcal{D}}$, should be even,
because the rectangular domain $D_{\gamma}$ contains the real axis.
Although we have no analytical proof that there exists a range of parameters
$\bm{\lambda}$ satisfying the condition \eqref{nozero},
we can verify that there exists such a range of $\bm{\lambda}$ by numerical
calculation (for small $M$ and $d_j$).
We have observed various sufficient conditions for the parameter range
satisfying \eqref{nozero}, with $M_{\I}M_{\II}\neq 0$ or $=0$ and
$\text{Im}\,a_i\neq 0$ or $=0$.
For example, for $M_{\II}=0$ case, the following parameter ranges seem
to be sufficient conditions:
\begin{align*}
  \cdot&\ \ M_{\I}=1,\ 0\leq d_1<2\text{Re}\,a_1-1,\ \text{$d_1$: even},
  \ \text{Re}\,a_1-\text{Re}\,a_2<\tfrac12(d_1+1),\\
  \cdot&\ \ M_{\I}=2,\ 0\leq d_1<d_2<2\text{Re}\,a_1-1,
  \ \text{$d_1$: even},\ \text{$d_2$: odd},
  \ \text{Re}\,a_1-\text{Re}\,a_2<\tfrac12(d_1+1),\\
  \cdot&\ \ M_{\I}=2,\ 0\leq d_1<d_2<2\text{Re}\,a_1-1,
  \ \text{$d_1$: even},\ d_2=d_1+1,\ \text{Re}\,a_1-\text{Re}\,a_2>d_2+2,\\
  \cdot&\ \ 0\leq d_1<d_2<\cdots<d_M<2\text{Re}\,a_1-1,
  \ (-1)^{d_j}=(-1)^{j-1}\ (1\leq j\leq M),
  \ \text{Re}\,a_1\ll\text{Re}\,a_2.
\end{align*}
In the following we assume that the condition \eqref{nozero} is satisfied.

If the deformed systems is well-defined, the general formula gives
the orthogonality of the eigenfunctions \cite{os27}:
\begin{equation}
  (\phi_{\mathcal{D}\,n},\phi_{\mathcal{D}\,m})
  =\prod_{j=1}^M(\mathcal{E}_n-\tilde{\mathcal{E}}_{d_j})\cdot
  h_n\delta_{nm}\ \ (n,m=0,1,2,\ldots).
  \label{orthoD}
\end{equation}
Namely, the orthogonality relations of the multi-indexed polynomials
$\check{P}_{\mathcal{D},n}(x;\bm{\lambda})$ are
\begin{align}
  &\int_{x_1}^{x_2}\!\!dx\,\psi_{\mathcal{D}}(x;\bm{\lambda})^2
  \check{P}_{\mathcal{D},n}(x;\bm{\lambda})
  \check{P}_{\mathcal{D},m}(x;\bm{\lambda})
  =h_{\mathcal{D},n}(\bm{\lambda})\delta_{nm}
  \ \ (n,m=0,1,2,\ldots),
  \label{orthocPDn}\\
  &h_{\mathcal{D},n}(\bm{\lambda})=h_n(\bm{\lambda})
  \prod_{j=1}^{M_{\I}}\bigl(\mathcal{E}_n(\bm{\lambda})
  -\tilde{\mathcal{E}}^{\I}_{d^{\I}_j}(\bm{\lambda})\bigr)\cdot
  \prod_{j=1}^{M_{\II}}\bigl(\mathcal{E}_n(\bm{\lambda})
  -\tilde{\mathcal{E}}^{\II}_{d^{\II}_j}(\bm{\lambda})\bigr).
  \label{hDn}
\end{align}
The multi-indexed orthogonal polynomial $P_{\mathcal{D},n}(\eta;\bm{\lambda})$
has $n$ zeros in the physical region $\eta\in\mathbb{R}$ ($\Leftrightarrow$
$\eta(x_1)<\eta<\eta(x_2)$), which interlace the $n+1$ zeros of
$P_{\mathcal{D},n+1}(\eta;\bm{\lambda})$ in the physical region,
and $\ell_{\mathcal{D}}$ zeros in the unphysical region
$\eta\in\mathbb{C}\backslash\mathbb{R}$.
These properties and \eqref{orthocPDn} can be verified by numerical calculation.

For the cases of type $\I$ only ($M_{\I}=M$, $M_{\II}=0$,
$\mathcal{D}=\{d_1,\ldots,d_M\}$), the expressions \eqref{cXiDdef} and
\eqref{cPDndef} are slightly simplified,
\begin{align}
  &\quad\text{W}_{\gamma}[\check{\xi}^{\I}_{d_1},\ldots,
  \check{\xi}^{\I}_{d_M}](x;\bm{\lambda})
  =\varphi_M(x)
  \check{\Xi}_{\mathcal{D}}(x;\bm{\lambda}),
  \label{cXiDIonly}\\
  &\quad\nu^{\I}(x;\bm{\lambda}+M\tilde{\bm{\delta}}^{\I})^{-1}
  \text{W}_{\gamma}[\check{\xi}^{\I}_{d_1},\ldots,\check{\xi}^{\I}_{d_M},
  \nu^{\I}\check{P}_n](x;\bm{\lambda})
  =\varphi_{M+1}(x)
  \check{P}_{\mathcal{D},n}(x;\bm{\lambda})\n[2pt]
  &=i^{\frac12M(M+1)}\left|
  \begin{array}{cccc}
  \check{\xi}^{\I}_{d_1}(x^{(M+1)}_1;\bm{\lambda})&\cdots&
  \check{\xi}^{\I}_{d_M}(x^{(M+1)}_1;\bm{\lambda})
  &r^{\I}_1(x^{(M+1)}_1)\check{P}_n(x^{_(M+1)}_1;\bm{\lambda})\\
  \check{\xi}^{\I}_{d_1}(x^{(M+1)}_2;\bm{\lambda})&\cdots&
  \check{\xi}^{\I}_{d_M}(x^{(M+1)}_2;\bm{\lambda})
  &r^{\I}_2(x^{(M+1)}_2)\check{P}_n(x^{(M+1)}_2;\bm{\lambda})\\
  \vdots&\cdots&\vdots&\vdots\\
  \check{\xi}^{\I}_{d_1}(x^{(M+1)}_{M+1};\bm{\lambda})&\cdots&
  \check{\xi}^{\I}_{d_M}(x^{(M+1)}_{M+1};\bm{\lambda})
  &r^{\I}_{M+1}(x^{(M+1)}_{M+1})\check{P}_n(x^{(M+1)}_{M+1};\bm{\lambda})\\
  \end{array}\right|,
  \label{cPDnIonly}
\end{align}
where $r^{\I}_j(x)=r^{\I}_j(x;\bm{\lambda},M+1)$.
The cases of type $\II$ only ($M_{\I}=0$, $M_{\II}=M$) are similar.

%%%%%%%%%%%%%%%%%%%%%%%%%%%%%%%%%%%%%%%%%%%%%%%
%                                             %
% 3.3 Shape invariance                        %
%                                             %
%%%%%%%%%%%%%%%%%%%%%%%%%%%%%%%%%%%%%%%%%%%%%%%
\subsection{Shape invariance}
\label{sec:shapeinv}

The shape invariance of the original system is inherited by the deformed
systems.
By the same argument of \cite{os27}, the Hamiltonian
$\mathcal{H}_{\mathcal{D}}(\bm{\lambda})$ is shape invariant:
\begin{equation}
  \mathcal{A}_{\mathcal{D}}(\bm{\lambda})
  \mathcal{A}_{\mathcal{D}}(\bm{\lambda})^{\dagger}
  =\kappa\mathcal{A}_{\mathcal{D}}(\bm{\lambda}+\bm{\delta})^{\dagger}
  \mathcal{A}_{\mathcal{D}}(\bm{\lambda}+\bm{\delta})
  +\mathcal{E}_1(\bm{\lambda}).
  \label{shapeinvD}
\end{equation}
As a consequence of the shape invariance, the actions of
$\mathcal{A}_{\mathcal{D}}(\bm{\lambda})$ and
$\mathcal{A}_{\mathcal{D}}(\bm{\lambda})^{\dagger}$ on the eigenfunctions
$\phi_{\mathcal{D}\,n}(x;\bm{\lambda})$ are
\begin{align}
  &\mathcal{A}_{\mathcal{D}}(\bm{\lambda})
  \phi_{\mathcal{D}\,n}(x;\bm{\lambda})
  =\kappa^{\frac{M}{2}}f_n(\bm{\lambda})
  \phi_{\mathcal{D}\,n-1}(x;\bm{\lambda}+\bm{\delta}),\n
  &\mathcal{A}_{\mathcal{D}}(\bm{\lambda})^{\dagger}
  \phi_{\mathcal{D}\,n-1}(x;\bm{\lambda}+\bm{\delta})
  =\kappa^{-\frac{M}{2}}b_{n-1}(\bm{\lambda})
  \phi_{\mathcal{D}\,n}(x;\bm{\lambda}).
  \label{ADphiDn=,ADdphiDn=}
\end{align}
The forward and backward shift operators are defined by
\begin{align}
  \mathcal{F}_{\mathcal{D}}(\bm{\lambda})&\eqdef
  \psi_{\mathcal{D}}\,(x;\bm{\lambda}+\bm{\delta})^{-1}\circ
  \mathcal{A}_{\mathcal{D}}(\bm{\lambda})\circ
  \psi_{\mathcal{D}}\,(x;\bm{\lambda})\n
  &=\frac{i}{\varphi(x)\check{\Xi}_{\mathcal{D}}(x;\bm{\lambda})}
  \Bigl(\check{\Xi}_{\mathcal{D}}(x+i\tfrac{\gamma}{2};
  \bm{\lambda}+\bm{\delta})e^{\frac{\gamma}{2}p}
  -\check{\Xi}_{\mathcal{D}}(x-i\tfrac{\gamma}{2};
  \bm{\lambda}+\bm{\delta})e^{-\frac{\gamma}{2}p}\Bigr),
  \label{calFD}\\
  \mathcal{B}_{\mathcal{D}}(\bm{\lambda})&\eqdef
  \psi_{\mathcal{D}}\,(x;\bm{\lambda})^{-1}\circ
  \mathcal{A}_{\mathcal{D}}(\bm{\lambda})^{\dagger}\circ
  \psi_{\mathcal{D}}\,(x;\bm{\lambda}+\bm{\delta})\n
  &=\frac{-i}{\check{\Xi}_{\mathcal{D}}(x;\bm{\lambda}+\bm{\delta})}
  \Bigl(V(x;\bm{\lambda}^{[M_{\I},M_{\II}]})
  \check{\Xi}_{\mathcal{D}}(x+i\tfrac{\gamma}{2};\bm{\lambda})
  e^{\frac{\gamma}{2}p}\n
  &\qquad\qquad\qquad\quad
  -V^*(x;\bm{\lambda}^{[M_{\I},M_{\II}]})
  \check{\Xi}_{\mathcal{D}}(x-i\tfrac{\gamma}{2};\bm{\lambda})
  e^{-\frac{\gamma}{2}p}\Bigr)\varphi(x),
  \label{calBD}
\end{align}
and their actions on $\check{P}_{\mathcal{D},n}(x;\bm{\lambda})$ are
\begin{align}
  &\mathcal{F}_{\mathcal{D}}(\bm{\lambda})
  \check{P}_{\mathcal{D},n}(x;\bm{\lambda})
  =f_n(\bm{\lambda})
  \check{P}_{\mathcal{D},n-1}(x;\bm{\lambda}+\bm{\delta}),\n
  &\mathcal{B}_{\mathcal{D}}(\bm{\lambda})
  \check{P}_{\mathcal{D},n-1}(x;\bm{\lambda}+\bm{\delta})
  =b_{n-1}(\bm{\lambda})
  \check{P}_{\mathcal{D},n}(x;\bm{\lambda}).\!
  \label{FDPDn=,BDPDn=}
\end{align}
The similarity transformed Hamiltonian is square root free:
\begin{align}
  \widetilde{\mathcal{H}}_{\mathcal{D}}(\bm{\lambda})
  &\eqdef\psi_{\mathcal{D}}(x;\bm{\lambda})^{-1}\circ
  \mathcal{H}_{\mathcal{D}}(\bm{\lambda})\circ
  \psi_{\mathcal{D}}(x;\bm{\lambda})
  =\mathcal{B}_{\mathcal{D}}(\bm{\lambda})
  \mathcal{F}_{\mathcal{D}}(\bm{\lambda})\n
  &=V(x;\bm{\lambda}^{[M_{\I},M_{\II}]})\,
  \frac{\check{\Xi}_{\mathcal{D}}(x+i\frac{\gamma}{2};\bm{\lambda})}
  {\check{\Xi}_{\mathcal{D}}(x-i\frac{\gamma}{2};\bm{\lambda})}
  \biggl(e^{\gamma p}
  -\frac{\check{\Xi}_{\mathcal{D}}(x-i\gamma;\bm{\lambda}+\bm{\delta})}
  {\check{\Xi}_{\mathcal{D}}(x;\bm{\lambda}+\bm{\delta})}\biggr)\n
  &\quad+V^*(x;\bm{\lambda}^{[M_{\I},M_{\II}]})\,
  \frac{\check{\Xi}_{\mathcal{D}}(x-i\frac{\gamma}{2};\bm{\lambda})}
  {\check{\Xi}_{\mathcal{D}}(x+i\frac{\gamma}{2};\bm{\lambda})}
  \biggl(e^{-\gamma p}
  -\frac{\check{\Xi}_{\mathcal{D}}(x+i\gamma;\bm{\lambda}+\bm{\delta})}
  {\check{\Xi}_{\mathcal{D}}(x;\bm{\lambda}+\bm{\delta})}\biggr),
  \label{tHD}
\end{align}
and the multi-indexed orthogonal polynomials
$\check{P}_{\mathcal{D},n}(x;\bm{\lambda})$ are its eigenpolynomials:
\begin{equation}
  \widetilde{\mathcal{H}}_{\mathcal{D}}(\bm{\lambda})
  \check{P}_{\mathcal{D},n}(x;\bm{\lambda})=\mathcal{E}_n(\bm{\lambda})
  \check{P}_{\mathcal{D},n}(x;\bm{\lambda}).
  \label{tHPDn=}
\end{equation}

The properties \eqref{dIM1=0}--\eqref{dIIM2=0} are also the consequences of
the shape invariance.
By \eqref{PD0=A.XiD}, the similar property holds for the denominator
polynomial $\check{\Xi}_{\mathcal{D}}(x;\bm{\lambda})$.
By the remark below \eqref{VD2}, the $M$-step deformed system labeled by
$\mathcal{D}$ with $0$ is equivalent to the $(M-1)$-step deformed
system labeled by $\mathcal{D}'$ with shifted parameters
$\bm{\lambda}+\tilde{\bm{\delta}}$.

%%%%%%%%%%%%%%%%%%%%%%%%%%%%%%%%%%%%%%%%%%%%%%%%%%%%%
%                                                   %
% 3.4 Limit from the Wilson system                  %
%                                                   %
%%%%%%%%%%%%%%%%%%%%%%%%%%%%%%%%%%%%%%%%%%%%%%%%%%%%%
\subsection{Limit from the Wilson system}
\label{sec:limWtocH}

The continuous Hahn polynomial can be obtained from the Wilson polynomial
\cite{kls}.
The potential function $V(x;\bm{\lambda})$, the energy eigenvalue
$\mathcal{E}_n(\bm{\lambda})$, the sinusoidal coordinate $\eta(x)$ and the
eigenfunctions $\phi_n(x;\bm{\lambda})$ of the Wilson system are \cite{os27}
\begin{align}
  &\bm{\lambda}=(a_1,a_2,a_3,a_4),\ \ \text{Re}\,a_i>0,
  \ \ \{a_1^*,a_2^*,a_3^*,a_4^*\}=\{a_1,a_2,a_3,a_4\}\ (\text{as a set}),
  \ \ b_1=\sum_{j=1}^4a_j,\n[-2pt]
  &V(x;\bm{\lambda})=\frac{\prod_{j=1}^4(a_j+ix)}{2ix(2ix+1)},\quad
  \mathcal{E}_n(\bm{\lambda})=n(n+b_1-1),\quad\eta(x)=x^2,\\
  &\phi_n(x;\bm{\lambda})=\phi_0(x;\bm{\lambda})\check{P}_n(x;\bm{\lambda}),
  \quad\phi_0(x;\bm{\lambda})=\sqrt{
  \frac{\prod_{j=1}^4\Gamma(a_j+ix)\Gamma(a_j-ix)}{\Gamma(2ix)\Gamma(-2ix)}},\\
  &\check{P}_n(x;\bm{\lambda})=P_n\bigl(\eta(x);\bm{\lambda}\bigr)
  =W_n\bigl(\eta(x);a_1,a_2,a_3,a_4\bigr)\n
  &\phantom{\check{P}_n(x;\bm{\lambda})}
  =(a_1+a_2,a_1+a_3,a_1+a_4)_n\cdot{}_4F_3\Bigl(
  \genfrac{}{}{0pt}{}{-n,\,n+b_1-1,\,a_1+ix,\,a_1-ix}
  {a_1+a_2,\,a_1+a_3,\,a_1+a_4}\Bigm|1\Bigr),
\end{align}
where $W_n(\eta;a_1,a_2,a_3,a_4)$ is the Wilson polynomial of degree $n$
in $\eta$ \cite{kls}.

Let us consider the following limit:
\begin{equation}
  x^{\text{W}}=x+t,\quad
  \bm{\lambda}^{\text{W}}=(a_1-it,a_1^*+it,a_2-it,a_2^*+it),\quad
  t\to\infty.
  \label{W->cH}
\end{equation}
Here the superscript $\text{W}$ indicates the quantities of the Wilson system,
and $x$, $a_1$ and $a_2$ are quantities of the continuous Hahn system.
The physical range of $x^{\text{W}}$ ($0\leq x^{\text{W}}<\infty$) gives
the physical range of $x$ ($-\infty<x<\infty$).
The continuous Hahn polynomial is obtained from the Wilson polynomial
\cite{kls},
\begin{equation}
  \lim_{t\to\infty}\frac{1}{(-2t)^n\,n!}\,
  \check{P}^{\text{W}}_n(x^{\text{W}};\bm{\lambda}^{\text{W}})
  =\check{P}_n(x;\bm{\lambda}).
\end{equation}
Note that $\check{P}^{\text{W}}_n(x^{\text{W}};\bm{\lambda}^{\text{W}})$ is
a polynomial in $\eta^{\text{W}}(x^{\text{W}})=(x^{\text{W}})^2$ and
$\check{P}_n(x;\bm{\lambda})$ is a polynomial in $\eta(x)=x$.
Other quantities are also obtained:
\begin{align}
  &\lim_{t\to\infty}V^{\text{W}}(x^{\text{W}};\bm{\lambda}^{\text{W}})
  =V(x;\bm{\lambda}),\quad
  \lim_{t\to\infty}\mathcal{E}^{\text{W}}_n(\bm{\lambda}^{\text{W}})
  =\mathcal{E}_n(\bm{\lambda}),\n
  &\lim_{t\to\infty}\frac{e^{\frac{\pi}{2}(\text{Im}\,(a_1+a_2)+2t)}}
  {\sqrt{2\pi(2t)^{b_1-1}}}\,
  \phi_0^{\text{W}}(x^{\text{W}};\bm{\lambda}^{\text{W}})
  =\phi_0(x;\bm{\lambda}).
\end{align}
The continuous Hahn system is obtained from the Wilson system by the limit
\eqref{W->cH}.

Next let us consider the deformed case.
We can show that the denominator polynomial
$\check{\Xi}_{\mathcal{D}}(x;\bm{\lambda})$ and the multi-indexed polynomials
$\check{P}_{\mathcal{D},n}(x;\bm{\lambda})$ of the continuous Hahn type
are obtained from those of the Wilson type by the same limit \eqref{W->cH}:
\begin{align}
  \lim_{t\to\infty}\frac{(-1)^{\frac12M(M-1)}}
  {(-2t)^{\ell_{\mathcal{D}}}\prod_{j=1}^Md_j!}\,
  \check{\Xi}^{\text{W}}_{\mathcal{D}}(x^{\text{W}};\bm{\lambda}^{\text{W}})
  &=\check{\Xi}_{\mathcal{D}}(x;\bm{\lambda}),
  \label{W->cH:XiD}\\
  \lim_{t\to\infty}\frac{(-1)^{\frac12M(M-1)}}
  {(-2t)^{\ell_{\mathcal{D}}+n}n!\prod_{j=1}^Md_j!}\,
  \check{P}^{\text{W}}_{\mathcal{D},n}(x^{\text{W}};\bm{\lambda}^{\text{W}})
  &=\check{P}_{\mathcal{D},n}(x;\bm{\lambda}),
\end{align}
where explicit forms of $\check{\Xi}^{\text{W}}_{\mathcal{D}}$ and
$\check{P}^{\text{W}}_{\mathcal{D},n}$ are found in \cite{os27}.
We remark that the denominator polynomial
$\check{\Xi}_{\mathcal{D}}(x;\bm{\lambda})$ is obtained as
\eqref{W->cH:XiD} algebraically, but the condition \eqref{nozero} is not
inherited from that of the Wilson type in general.
Therefore the limit \eqref{W->cH} of the deformed Wilson systems do not
give the deformed continuous Hahn systems in general.

%%%%%%%%%%%%%%%%%%%%%%%%%%%%%%%%%%%%%%%%%%%%%%%%%%%%%%%%%%%%%%%
%                                                             %
%  4. New Exactly Solvable idQM Systems and                   %
%     Multi-indexed Meixner-Pollaczek Polynomials             %
%                                                             %
%%%%%%%%%%%%%%%%%%%%%%%%%%%%%%%%%%%%%%%%%%%%%%%%%%%%%%%%%%%%%%%
\section{New Exactly Solvable idQM Systems and Multi-indexed
Meixner-Pollaczek Polynomials}
\label{sec:newidQMMP}

In this section we deform the Meixner-Pollaczek system.
Since the method is the same as in \S\,\ref{sec:newidQMcH}, we present results
briefly.
The eigenfunctions of the deformed systems are described by the case-(1)
multi-indexed Meixner-Pollaczek polynomials.

%%%%%%%%%%%%%%%%%%%%%%%%%%%%%%%%%%%%%%%%%%%%%%%
%                                             %
% 4.1 Original Meixner-Pollaczek system       %
%                                             %
%%%%%%%%%%%%%%%%%%%%%%%%%%%%%%%%%%%%%%%%%%%%%%%
\subsection{Original Meixner-Pollaczek system}
\label{sec:MP}

The Meixner-Pollaczek system is the idQM system with \eqref{x1x2gamma} and
a set of parameters $\bm{\lambda}$ is
\begin{equation}
  \bm{\lambda}=(a,\phi),\quad a>0,\quad 0<\phi<\pi.
\end{equation}
The fundamental data are the following \cite{os13}:
\begin{align}
  &V(x;\bm{\lambda})=e^{i(\frac{\pi}{2}-\phi)}(a+ix),\quad
  \eta(x)=x,\ \ \varphi(x)=1,
  \ \ \mathcal{E}_n(\bm{\lambda})=2n\sin\phi,\\
  &\phi_n(x;\bm{\lambda})
  =\phi_0(x;\bm{\lambda})\check{P}_n(x;\bm{\lambda}),\quad
  \phi_0(x;\bm{\lambda})=e^{(\phi-\frac{\pi}{2})x}
  \sqrt{\Gamma(a+ix)\Gamma(a-ix)},\\
  &\check{P}_n(x;\bm{\lambda})=P_n\bigl(\eta(x);\bm{\lambda}\bigr)
  =P^{(a)}_n\bigl(\eta(x);\phi\bigr)
  =\frac{(2a)_n}{n!}e^{in\phi}
  {}_2F_1\Bigl(\genfrac{}{}{0pt}{}{-n,\,a+ix}
  {2a}\Bigm|1-e^{-2i\phi}\Bigr),
  \label{PnMP}\\
  &h_n(\bm{\lambda})=
  2\pi \frac{\Gamma(n+2a)}{n!\,(2\sin\phi)^{2a}},\quad
  c_n(\bm{\lambda})=\frac{(2\sin\phi)^n}{n!},\\
  &\bm{\delta}=(\tfrac12,0),\ \ \kappa=1,
  \ \ f_n(\bm{\lambda})=2\sin\phi,\ \ b_{n-1}(\bm{\lambda})=n.
\end{align}
Here $P^{(a)}_n(\eta;\phi)$ in \eqref{PnMP} is the Meixner-Pollaczek
polynomial of degree $n$ in $\eta$ \cite{kls}.
%Note that $\phi^*_0(x;\bm{\lambda})=\phi_0(x;\bm{\lambda})$
%and $\check{P}^*_n(x;\bm{\lambda})=\check{P}_n(x;\bm{\lambda})$.
%It is not necessary to distinguish $\check{P}_n$ and $P_n$ since $\eta(x)=x$,
%but we will use both notations to compare with other cases in \cite{os27}.

%%%%%%%%%%%%%%%%%%%%%%%%%%%%%%%%%%%%%%%%%%%%%%%
%                                             %
% 4.2 Virtual state wavefunctions             %
%                                             %
%%%%%%%%%%%%%%%%%%%%%%%%%%%%%%%%%%%%%%%%%%%%%%%
\subsection{Virtual state wavefunctions}
\label{sec:cH:vs}

Let us introduce a twist operation $\mathfrak{t}$,
\begin{equation}
  \mathfrak{t}(\bm{\lambda})\eqdef(1-a,\phi),\quad
  \tilde{\bm{\delta}}\eqdef(-\tfrac12,0),
\end{equation}
which is an involution $\mathfrak{t}^2=\text{id}$ and satisfies
$\mathfrak{t}(\bm{\lambda}+\beta\bm{\delta})=\mathfrak{t}(\bm{\lambda})
+\beta\tilde{\bm{\delta}}$ ($\beta\in\mathbb{R}$).
The potential function $V(x;\bm{\lambda})$ satisfies \eqref{propV'} with
\begin{equation}
  \alpha(\bm{\lambda})=1,\quad
  \alpha^{\prime}(\bm{\lambda})=2(1-2a)\sin\phi.
\end{equation}
In the following, we assume $a>\frac12$, which gives $\alpha'(\bm{\lambda})<0$.
The virtual state wavefunctions
$\tilde{\phi}_{\text{v}}(x;\bm{\lambda})$
($\text{v}\in\mathcal{V}\subset\mathbb{Z}_{\geq 0}$) are defined by
\begin{align}
  &\tilde{\phi}_{\text{v}}(x;\bm{\lambda})\eqdef
  \phi_{\text{v}}\bigl(x;\mathfrak{t}(\bm{\lambda})\bigr)
  =\tilde{\phi}_0(x;\bm{\lambda})
  \check{\xi}_{\text{v}}(x;\bm{\lambda}),
  \ \ \tilde{\phi}_0(x;\bm{\lambda})\eqdef
  \phi_0\bigl(x;\mathfrak{t}(\bm{\lambda})\bigr),\n
  &\check{\xi}_{\text{v}}(x;\bm{\lambda})\eqdef
  \xi_{\text{v}}\bigl(\eta(x);\bm{\lambda}\bigr)\eqdef
  \check{P}_{\text{v}}\bigl(x;\mathfrak{t}(\bm{\lambda})\bigr)
  =P_{\text{v}}\bigl(\eta(x);\mathfrak{t}(\bm{\lambda})\bigr)
  \ \ (\text{v}\in\mathcal{V}),
\end{align}
which satisfy the Schr\"odinger equation
$\mathcal{H}(\bm{\lambda})\tilde{\phi}_{\text{v}}(x;\bm{\lambda})
=\tilde{\mathcal{E}}_{\text{v}}(\bm{\lambda})
\tilde{\phi}_{\text{v}}(x;\bm{\lambda})$.
The virtual state polynomial $\xi_{\text{v}}(\eta;\bm{\lambda})$
is a polynomial of degree $\text{v}$ in $\eta$, and the virtual energy
$\tilde{\mathcal{E}}_{\text{v}}(\bm{\lambda})$ is
\begin{equation}
  \tilde{\mathcal{E}}_{\text{v}}(\bm{\lambda})
  =2(\text{v}+1-2a)\sin\phi,
\end{equation}
which is negative for $2a>\text{v}+1$.
We choose $\mathcal{V}$ as
\begin{equation}
  \mathcal{V}=\bigl\{0,1,2,\ldots,[2a-1]'\bigr\}.
  \label{MP:vrange}
\end{equation}
%where $[x]'$ denotes the greatest integer not equal or exceeding $x$.
%Although we have included 0 in $\mathcal{V}$, the Darboux transformations with
%the label 0 virtual state do not give essentially new systems, see the end of
%\S\,\ref{sec:shapeinv}.

%%%%%%%%%%%%%%%%%%%%%%%%%%%%%%%%%%%%%%%%%%%%%%%
%                                             %
% 4.3 New exactly solvable systems            %
%                                             %
%%%%%%%%%%%%%%%%%%%%%%%%%%%%%%%%%%%%%%%%%%%%%%%
\subsection{New exactly solvable systems}
\label{sec:MP:newsys}

Isospectral deformations of the Meixner-Pollaczek system are obtained by
the multi-step Darboux transformations with the virtual state wavefunctions
as seed solutions.
The deformed systems are labeled by
$\mathcal{D}=\{d_1,\ldots,d_M\}$ ($d_j\in\mathcal{V}$ : mutually distinct).

Let us define the following functions:
\begin{equation}
  \nu(x;\bm{\lambda})\eqdef
  \frac{\phi_0(x;\bm{\lambda})}{\tilde{\phi}_0(x;\bm{\lambda})},
  \ \ r_j^{\I}(x^{(M)}_j;\bm{\lambda},M)\eqdef
  \frac{\nu(x^{(M)}_j;\bm{\lambda})}
  {\nu\bigl(x;\bm{\lambda}+(M-1)\tilde{\bm{\delta}}\bigr)}
  \ \ (j=1,2,\ldots,M),
\end{equation}
whose explicit form is
\begin{equation}
  r_j(x^{(M)}_j;\bm{\lambda},M)
  =(-1)^{j-1}i^{1-M}(a-\tfrac{M-1}{2}+ix)_{j-1}
  (a-\tfrac{M-1}{2}-ix)_{M-j}.
\end{equation}
The denominator polynomial and the multi-indexed polynomial are defined by
\eqref{XiP_poly} and
\begin{align}
  &\quad\text{W}_{\gamma}[\check{\xi}_{d_1},\ldots,
  \check{\xi}_{d_M}](x;\bm{\lambda})
  =\varphi_M(x)
  \check{\Xi}_{\mathcal{D}}(x;\bm{\lambda}),
  \label{MP:cXiDdef}\\
  &\quad\nu(x;\bm{\lambda}+M\tilde{\bm{\delta}})^{-1}
  \text{W}_{\gamma}[\check{\xi}_{d_1},\ldots,\check{\xi}_{d_M},
  \nu\check{P}_n](x;\bm{\lambda})
  =\varphi_{M+1}(x)
  \check{P}_{\mathcal{D},n}(x;\bm{\lambda})\n[2pt]
  &=i^{\frac12M(M+1)}\left|
  \begin{array}{cccc}
  \check{\xi}_{d_1}(x^{(M+1)}_1;\bm{\lambda})&\cdots&
  \check{\xi}_{d_M}(x^{(M+1)}_1;\bm{\lambda})
  &r_1(x^{(M+1)}_1)\check{P}_n(x^{_(M+1)}_1;\bm{\lambda})\\
  \check{\xi}_{d_1}(x^{(M+1)}_2;\bm{\lambda})&\cdots&
  \check{\xi}_{d_M}(x^{(M+1)}_2;\bm{\lambda})
  &r_2(x^{(M+1)}_2)\check{P}_n(x^{(M+1)}_2;\bm{\lambda})\\
  \vdots&\cdots&\vdots&\vdots\\
  \check{\xi}_{d_1}(x^{(M+1)}_{M+1};\bm{\lambda})&\cdots&
  \check{\xi}_{d_M}(x^{(M+1)}_{M+1};\bm{\lambda})
  &r_{M+1}(x^{(M+1)}_{M+1})\check{P}_n(x^{(M+1)}_{M+1};\bm{\lambda})\\
  \end{array}\right|,
  \label{MP:cPDndef}
\end{align}
where $r_j(x)=r_j(x;\bm{\lambda},M+1)$ and $\varphi_M(x)=1$.
%(see \eqref{cXiDIonly}--\eqref{cPDnIonly})
Their degrees are $\ell_{\mathcal{D}}$ and $\ell_{\mathcal{D}}+n$,
respectively (we assume $c_{\mathcal{D}}^{\Xi}(\bm{\lambda})\neq 0$ and
$c_{\mathcal{D},n}^{P}(\bm{\lambda})\neq 0$,
see \eqref{MP:cXiD}--\eqref{MP:cPDn}).
Here $\ell_{\mathcal{D}}$ is \eqref{lD} with $M_{\II}=0$.
These $\check{\Xi}_{\mathcal{D}}(x;\bm{\lambda})$ and
$\check{P}_{\mathcal{D},n}(x;\bm{\lambda})$ are `real',
$\check{\Xi}^*_{\mathcal{D}}(x;\bm{\lambda})
=\check{\Xi}_{\mathcal{D}}(x;\bm{\lambda})$ and
$\check{P}^*_{\mathcal{D},n}(x;\bm{\lambda})
=\check{P}_{\mathcal{D},n}(x;\bm{\lambda})$.
Then, the Casoratians 
$\text{W}_{\gamma}[\tilde{\phi}_{d_1},\ldots,\tilde{\phi}_{d_M}](x)$ and
$\text{W}_{\gamma}[\tilde{\phi}_{d_1},\ldots,\tilde{\phi}_{d_M},\phi_n](x)$
are expressed as
\begin{align}
  &\quad\text{W}_{\gamma}[\tilde{\phi}_{d_1},\ldots,\tilde{\phi}_{d_M}]
  (x;\bm{\lambda})=\prod_{j=1}^M\phi_0\bigl(x^{(M)}_j;\bm{\lambda}\bigr)\cdot
  \text{W}_{\gamma}\bigl[\check{\xi}_{d_1},\ldots,\check{\xi}_{d_M}\bigr]
  (x;\bm{\lambda})\n
  &=\prod_{j=1}^M\phi_0\bigl(x^{(M)}_j;\bm{\lambda}\bigr)
  \times\varphi_M(x)\check{\Xi}_{\mathcal{D}}(x;\bm{\lambda}),\\
  &\quad\text{W}_{\gamma}[\tilde{\phi}_{d_1},\ldots,\tilde{\phi}_{d_M},\phi_n]
  (x;\bm{\lambda})
  =\prod_{j=1}^{M+1}\phi_0\bigl(x^{(M+1)}_j;\bm{\lambda}\bigr)\cdot
  \text{W}_{\gamma}\bigl[\check{\xi}_{d_1},\ldots,\check{\xi}_{d_M},
  \nu\check{P}_n\bigr](x;\bm{\lambda})\n
  &=\prod_{j=1}^{M+1}\phi_0\bigl(x^{(M+1)}_j;\bm{\lambda}\bigr)
  \times\nu\bigl(x;\bm{\lambda}+M\tilde{\bm{\delta}}\bigr)
  \varphi_{M+1}(x)\check{P}_{\mathcal{D},n}(x;\bm{\lambda}).
\end{align}
The eigenfunction $\phi_{\mathcal{D}\,n}(x;\bm{\lambda})$ \eqref{phiDn} is
rewritten as \eqref{phiDn2} and $\psi_{\mathcal{D}}(x;\bm{\lambda})$ is
\begin{equation}
  \psi_{\mathcal{D}}(x;\bm{\lambda})\eqdef
  \frac{\phi_0(x;\bm{\lambda}+M\tilde{\bm{\delta}})}
  {\sqrt{\check{\Xi}_{\mathcal{D}}(x-i\frac{\gamma}{2};\bm{\lambda})
  \check{\Xi}_{\mathcal{D}}(x+i\frac{\gamma}{2};\bm{\lambda})}}.
  \label{MPpsiD}
\end{equation}
The ground state wavefunction $\phi_{\mathcal{D}\,0}$ is annihilated by
$\mathcal{A}_{\mathcal{D}}$,
$\mathcal{A}_{\mathcal{D}}(\bm{\lambda})
\phi_{\mathcal{D}\,0}(x;\bm{\lambda})=0$.
The lowest degree multi-indexed orthogonal polynomial
$\check{P}_{\mathcal{D},0}(x;\bm{\lambda})$ is proportional to
$\check{\Xi}_{\mathcal{D}}(x;\bm{\lambda}+\bm{\delta})$,
see \eqref{MP:PD0=A.XiD}.
The potential function $V_{\mathcal{D}}(x)$ \eqref{VD} is expressed as
\begin{equation}
  V_{\mathcal{D}}(x;\bm{\lambda})
  =V(x;\bm{\lambda}+M\tilde{\bm{\delta}})\,
  \frac{\check{\Xi}_{\mathcal{D}}(x+i\frac{\gamma}{2};\bm{\lambda})}
  {\check{\Xi}_{\mathcal{D}}(x-i\frac{\gamma}{2};\bm{\lambda})}
  \frac{\check{\Xi}_{\mathcal{D}}(x-i\gamma;\bm{\lambda}+\bm{\delta})}
  {\check{\Xi}_{\mathcal{D}}(x;\bm{\lambda}+\bm{\delta})}.
  \label{MPVD2}
\end{equation}

To check the regularity and hermiticity of
$\mathcal{H}_{\mathcal{D}}(\bm{\lambda})$,
let us consider the function $g(x)$,
\begin{align}
  g(x)&\eqdef V(x+i\tfrac{\gamma}{2};\bm{\lambda}+M\tilde{\bm{\delta}})
  \phi_0(x+i\tfrac{\gamma}{2};\bm{\lambda}+M\tilde{\bm{\delta}})^2\n
  &=e^{2(\phi-\frac{\pi}{2})x}\,
  \Gamma(a-\tfrac12M+\tfrac12+ix)\Gamma(a-\tfrac12M+\tfrac12-ix).
\end{align}
Asymptotic behavior of $g(x)$ at $x\sim\pm\infty$ is
$g(x)\sim 2\pi|x|^{2a-M}e^{2(\phi-\frac{\pi}{2})x-\pi|x|}$
(for $x\in\mathbb{R}$).
The necessary and sufficient condition for $g(x)$ to have no poles in the
rectangular domain $D_{\gamma}$ is $a-\frac12M>0$.
This condition is automatically satisfied because of \eqref{MP:vrange}.
By the same argument as \S\,\ref{sec:newsys},
the deformed Hamiltonian $\mathcal{H}_{\mathcal{D}}(\bm{\lambda})$ is
well-defined and hermitian, if the condition \eqref{nozero} is satisfied.
To satisfy the condition \eqref{nozero}, the degree of
$\Xi_{\mathcal{D}}(\eta;\bm{\lambda})$, $\ell_{\mathcal{D}}$, should be even.
Although we have no analytical proof that there exists a range of parameters
$\bm{\lambda}$ satisfying the condition \eqref{nozero}, numerical calculation
(for small $M$ and $d_j$) suggests the following conjecture.
\begin{conj}
Let $d_j$'s be $0\leq d_1<d_2<\cdots<d_M<2a-1$.
Then, the necessary and sufficient condition for the condition \eqref{nozero}
is $(-1)^{d_j}=(-1)^{j-1}$ $(j=1,2,\ldots,M)$.
\end{conj}
In the following we assume that the condition \eqref{nozero} is satisfied.

If the deformed systems is well-defined, the eigenfunctions are orthogonal.
Namely, the orthogonality relations of the multi-indexed polynomials
$\check{P}_{\mathcal{D},n}(x;\bm{\lambda})$ are \eqref{orthocPDn} with
\begin{equation}
  h_{\mathcal{D},n}(\bm{\lambda})=h_n(\bm{\lambda})
  \prod_{j=1}^{M}\bigl(\mathcal{E}_n(\bm{\lambda})
  -\tilde{\mathcal{E}}_{d_j}(\bm{\lambda})\bigr).
  \label{MP:hDn}
\end{equation}
The multi-indexed orthogonal polynomial $P_{\mathcal{D},n}(\eta;\bm{\lambda})$
has $n$ zeros in the physical region $\eta\in\mathbb{R}$ ($\Leftrightarrow$
$\eta(x_1)<\eta<\eta(x_2)$), which interlace the $n+1$ zeros of
$P_{\mathcal{D},n+1}(\eta;\bm{\lambda})$ in the physical region,
and $\ell_{\mathcal{D}}$ zeros in the unphysical region
$\eta\in\mathbb{C}\backslash\mathbb{R}$.
These properties and \eqref{orthocPDn} can be verified by numerical
calculation.

The shape invariance of the original system is inherited by the deformed
systems.
The properties \eqref{shapeinvD}--\eqref{tHPDn=} (with the replacement
$\bm{\lambda}^{[M_{\I},M_{\II}]}\to\bm{\lambda}+M\tilde{\bm{\delta}}$) hold.
%\eqref{shapeinvD}, \eqref{ADphiDn=,ADdphiDn=}, \eqref{calFD}, \eqref{calBD},
%\eqref{FDPDn=,BDPDn=}, \eqref{tHD}, \eqref{tHPDn=}
The property \eqref{MP:dM=0} implies that
the $M$-step deformed system labeled by
$\mathcal{D}$ with $0$ is equivalent to the $(M-1)$-step deformed
system labeled by $\mathcal{D}'$ with shifted parameters
$\bm{\lambda}+\tilde{\bm{\delta}}$.

%%%%%%%%%%%%%%%%%%%%%%%%%%%%%%%%%%%%%%%%%%%%%%%%%%%%%
%                                                   %
% 4.4 Limit from the continuous Hahn system         %
%                                                   %
%%%%%%%%%%%%%%%%%%%%%%%%%%%%%%%%%%%%%%%%%%%%%%%%%%%%%
\subsection{Limit from the continuous Hahn system}
\label{sec:limcHtoMP}

The Meixner-Pollaczek polynomial can be obtained from the continuous Hahn
polynomial \cite{kls}.
Let us consider the following limit:
\begin{equation}
  x^{\text{cH}}=x+\frac{t}{\tan\phi}\,,\quad
  \bm{\lambda}^{\text{cH}}=\Bigl(a-i\frac{t}{\tan\phi}\,,\,t\Bigr),\quad
  t\to\infty.
  \label{cH->MP}
\end{equation}
Here the superscript $\text{cH}$ indicates the quantities of the
continuous Hahn system.

Under this limit \eqref{cH->MP}, the Meixner-Pollaczek polynomial is
obtained from the continuous Hahn polynomial \cite{kls},
\begin{equation}
  \lim_{t\to\infty}\Bigl(\frac{\sin\phi}{t}\Bigr)^n
  \check{P}^{\text{cH}}_n(x^{\text{cH}};\bm{\lambda}^{\text{cH}})
  =\check{P}_n(x;\bm{\lambda}).
\end{equation}
Other quantities are also obtained:
\begin{align}
  &\lim_{t\to\infty}\frac{\sin\phi}{t}\,
  V^{\text{cH}}(x^{\text{cH}};\bm{\lambda}^{\text{cH}})
  =V(x;\bm{\lambda}),\quad
  \lim_{t\to\infty}\frac{\sin\phi}{t}\,
  \mathcal{E}^{\text{cH}}_n(\bm{\lambda}^{\text{cH}})
  =\mathcal{E}_n(\bm{\lambda}),\n
  &\lim_{t\to\infty}\Bigl(\frac{\sin\phi}{t}\Bigr)^{t-\frac12}\,
  \frac{e^{\frac{t}{\tan\phi}(\frac{\pi}{2}-\phi)+t}}{\sqrt{2\pi}}\,
  \phi_0^{\text{cH}}(x^{\text{cH}};\bm{\lambda}^{\text{cH}})
  =\phi_0(x;\bm{\lambda}).
\end{align}
The Meixner-Pollaczek system is obtained from the continuous Hahn system
by the limit \eqref{cH->MP}.

Next let us consider the deformed case.
Under the limit \eqref{cH->MP}, the type $\I$ twist operation of the continuous
Hahn system $\mathfrak{t}^{\text{cH}\,\I}$ reduces to the twist operation of
the Meixner-Pollaczek system $\mathfrak{t}$, but type $\II$
$\mathfrak{t}^{\text{cH}\,\II}$ does not have a good limit.
Hence we consider the deformed continuous Hahn systems with $M_{\II}=0$.
We can show that the denominator polynomial
$\check{\Xi}_{\mathcal{D}}(x;\bm{\lambda})$ and the multi-indexed polynomials
$\check{P}_{\mathcal{D},n}(x;\bm{\lambda})$ of the Meixner-Pollaczek type
are obtained from those of the continuous Hahn type (with type $\I$ only)
by the limit \eqref{cH->MP}
\begin{align}
  \lim_{t\to\infty}\Bigl(\frac{\sin\phi}{t}\Bigr)^{\sum_{j=1}^Md_j}\,
  \check{\Xi}^{\text{cH}}_{\mathcal{D}}(x^{\text{cH}};\bm{\lambda}^{\text{cH}})
  &=\check{\Xi}_{\mathcal{D}}(x;\bm{\lambda}),
  \label{cH->MP:XiD}\\
  \lim_{t\to\infty}\Bigl(\frac{\sin\phi}{t}\Bigr)^{\sum_{j=1}^Md_j+n}\,
  \check{P}^{\text{cH}}_{\mathcal{D},n}(x^{\text{cH}};\bm{\lambda}^{\text{cH}})
  &=\check{P}_{\mathcal{D},n}(x;\bm{\lambda}).
\end{align}
We conjecture that the condition \eqref{nozero} of the Meixner-Pollaczek
system is obtained from that of the continuous Hahn system.
If this is true, the deformed Meixner-Pollaczek systems are obtained from
the deformed continuous Hahn systems (with type $\I$ only) by the limit
\eqref{cH->MP}.

%%%%%%%%%%%%%%%%%%%%%%%%%%%%%%%%%%%%%%%%%%%%%%%%%%%%%
%                                                   %
% 4.5 Limit to the harmonic oscillator              %
%                                                   %
%%%%%%%%%%%%%%%%%%%%%%%%%%%%%%%%%%%%%%%%%%%%%%%%%%%%%
\subsection{Limit to the harmonic oscillator}
\label{sec:MP->HO}

The harmonic oscillator is an ordinary quantum mechanical system
%defined on the whole line
and its eigenfunctions are the following:
\begin{align}
  &H=p^2+x^2-1,\quad\mathcal{E}_n=2n,\quad\eta(x)=x,\quad
  -\infty<x<\infty,\\
  &\phi_n(x)=\phi_0(x)\check{P}_n(x),\quad\phi_0(x)=e^{-\frac12x^2},\\
  &\check{P}_n(x)=P_n\bigl(\eta(x)\bigr)=H_n\bigl(\eta(x)\bigr)
  =(2x)^n\cdot{}_2F_0\Bigl(
  \genfrac{}{}{0pt}{}{-\tfrac{n}{2},\,-\tfrac{n-1}{2}}
  {-}\Bigm|-\frac{1}{x^2}\Bigr),
\end{align}
where $H_n(\eta)$ is the Hermite polynomial of degree $n$ in $\eta$ \cite{kls}.
%So we call the harmonic oscillator as the Hermite system.

Let us consider the following limit of the Meixner-Pollaczek system:
\begin{equation}
  x^{\text{MP}}=\sqrt{t}\,x,\quad
  \bm{\lambda}^{\text{MP}}=(t,\tfrac{\pi}{2}),\quad
  t\to\infty.
  \label{MP->HO}
\end{equation}
Under this limit, the Meixner-Pollaczek system reduces to the harmonic
oscillator:
\begin{align}
  &\lim_{t\to\infty}H^{\text{MP}}(\bm{\lambda}^{\text{MP}})=H,\quad
  \lim_{t\to\infty}\mathcal{E}^{\text{MP}}_n(\bm{\lambda}^{\text{MP}})
  =\mathcal{E}_n,\\
  &\lim_{t\to\infty}\frac{e^t}{\sqrt{2\pi}\,t^{t-\frac12}}\,
  \phi^{\text{MP}}_0(x^{\text{MP}};\bm{\lambda}^{\text{MP}})
  =\phi_0(x),\quad
  \lim_{t\to\infty}\frac{n!}{t^{\frac{n}{2}}}\,
  \check{P}^{\text{MP}}_n(x^{\text{MP}};\bm{\lambda}^{\text{MP}})
  =\check{P}_n(x).
\end{align}
We remark that the Meixner-Pollaczek polynomial $P^{(a)}_n(x;\phi)$ with
any $\phi$ reduces to the Hermite polynomial as \cite{kls}
\begin{equation}
  \lim_{t\to\infty}\frac{n!}{t^{\frac{n}{2}}}\,
  P^{(t)}_n\Bigl(\frac{\sqrt{t}\,x-t\cos\phi}{\sin\phi};\phi\Bigr)
  =H_n(x),
\end{equation}
 but this limit does not lead to a good limit of the quantum system.

Next let us consider the deformed case.
There is no virtual state in the harmonic oscillator \cite{os29}.
Hence the limit \eqref{MP->HO} of the virtual state wavefunction of the
Meixner-Pollaczek system can not be a virtual state wavefunction.
In fact, the limit of $\tilde{\phi}^{\text{MP}}_{\text{v}}$ is
%(x^{\text{MP}};\bm{\lambda}^{\text{MP}})$ is
\begin{equation}
  \lim_{t\to\infty}\frac{t^{t-\frac12}e^{-t}}{\sqrt{2\pi}}\,
  \tilde{\phi}^{\text{MP}}_0(x^{\text{MP}};\bm{\lambda}^{\text{MP}})
  =e^{\frac12x^2},\quad
  \lim_{t\to\infty}\frac{\text{v}!}{t^{\frac{\text{v}}{2}}}\,
  \check{\xi}^{\text{MP}}_{\text{v}}(x^{\text{MP}};\bm{\lambda}^{\text{MP}})
  =i^{-\text{v}}H_{\text{v}}(ix).
\end{equation}
This is the pseudo virtual state wavefunction of the harmonic oscillator
$\tilde{\phi}_{\text{v}}(x)=i^{-\text{v}}\phi_{\text{v}}(ix)$ \cite{os29}.
The deformed harmonic oscillator system, which is obtained by the Darboux
transformations with $\tilde{\phi}_{\text{v}}$ ($\text{v}\in\mathcal{D}$)
as seed solutions, has energy eigenvalues $\mathcal{E}_n$ ($n=0,1,\ldots$)
and $\tilde{\mathcal{E}}_{d_j}$ ($j=1,\ldots,M$).
The eigenfunctions with $\mathcal{E}_n$ are obtained as the limit
of $\phi^{\text{MP}}_{\mathcal{D}\,n}$,
%$(x^{\text{MP}};\bm{\lambda}^{\text{MP}})$,
but those with $\tilde{\mathcal{E}}_{d_j}$ can not be obtained from the
eigenfunctions of the deformed Meixner-Pollaczek system.
In this sense the limit \eqref{MP->HO} of the deformed Meixner-Pollaczek
system is not a good limit.

%%%%%%%%%%%%%%%%%%%%%%%%%%%%%%%%%%%%%%%%%%%%%%%%%%%%%%%%%%%%%%%
%                                                             %
%  5. Summary and Comments                                    %
%                                                             %
%%%%%%%%%%%%%%%%%%%%%%%%%%%%%%%%%%%%%%%%%%%%%%%%%%%%%%%%%%%%%%%
\section{Summary and Comments}
\label{sec:summary}

The continuous Hahn and Meixner-Pollaczek idQM systems are exactly solvable
and their physical range of the coordinate is the whole real line.
We deform them by the multi-step Darboux transformations with the virtual
state wavefunctions as seed solutions, and obtain new exactly solvable
idQM systems and the case-(1) multi-indexed continuous Hahn and
Meixner-Pollaczek polynomials.
By this result, the construction of the multi-indexed polynomials in idQM
is essentially completed.
The remaining task is to study the properties of various multi-indexed
polynomials and to use them to investigate quantum mechanical systems.

The deformed quantum system labeled by an index set $\mathcal{D}$ may be
equivalent to another labeled by a different index set $\mathcal{D}'$ with
shifted parameters, which means that the corresponding two multi-indexed
orthogonal polynomials labeled by $\mathcal{D}$ and $\mathcal{D}'$ with
shifted parameters are proportional.
Such equivalence is studied for the case-(2) multi-indexed polynomials of
Hermite, Laguerre, Jacobi, Wilson and Askey-Wilson types \cite{os29,os30}
and for the case-(1) multi-indexed polynomials of Laguerre, Jacobi, Wilson
and Askey-Wilson types \cite{equiv_miop} (see also \cite{t13}).
The case-(1) multi-indexed continuous Hahn polynomials obtained in this
paper have equivalence in the same form as the (Askey-)Wilson cases
\cite{equiv_miop}, which is derived from the properties
\eqref{dIM1=0}--\eqref{dIIM2=0}.
The case-(1) multi-indexed Meixner-Pollaczek polynomials also have similar
equivalence derived from the property \eqref{MP:dM=0}.

The multi-indexed orthogonal polynomials do not satisfy the three term
recurrence relations, which are characterizations of the ordinary orthogonal
polynomials \cite{ismail}, because they are not ordinary orthogonal
polynomials. Instead, they satisfy the recurrence relations with more terms
%\cite{stz10,rrmiop,d14_2,mt14,rrmiop2,gkkm15,rrmiop3,rrmiop4,rrmiop5}.
\cite{stz10}--\cite{rrmiop5}.
The case-(1) multi-indexed continuous Hahn and Meixner-Pollaczek polynomials
satisfy such recurrence relations. The recurrence relations with constant
coefficients are related to the generalized closure relations \cite{rrmiop4},
which give the creation and annihilation operators of the deformed quantum
systems.
We will report these topics elsewhere \cite{rrmiop6}.

%%%%%%%%%%%%%%%%%%%%%%%%%%%%%%%%%%%%%%%%%%%%%%%%%%%%%%%%%%%%%%%
%                                                             %
%  Acknowledgments                                            %
%                                                             %
%%%%%%%%%%%%%%%%%%%%%%%%%%%%%%%%%%%%%%%%%%%%%%%%%%%%%%%%%%%%%%%
\section*{Acknowledgments}

%I thank R.\,Sasaki for discussion and useful comments on the manuscript.
This work was supported by JSPS KAKENHI Grant Numbers JP19K03667.

\bigskip
\appendix
%\renewcommand{\theequation}{\Alph{section}.\arabic{equation}}
%%%%%%%%%%%%%%%%%%%%%%%%%%%%%%%%%%%%%%%%%%%%%%%%%%%%%%%%%%%%%%%
%                                                             %
%  A. Some Properties of the Multi-indexed                    %
%     Continuous Hahn Polynomials                             %
%                                                             %
%%%%%%%%%%%%%%%%%%%%%%%%%%%%%%%%%%%%%%%%%%%%%%%%%%%%%%%%%%%%%%%
\section{Some Properties of the Multi-indexed Continuous Hahn Polynomials}
\label{app:prop}

We present some properties of the multi-indexed continuous Hahn polynomials.

\noindent
$\bullet$ coefficients of the highest degree terms:
\begin{align}
  &\Xi_{\mathcal{D}}(\eta;\bm{\lambda})
  =c_{\mathcal{D}}^{\Xi}(\bm{\lambda})\eta^{\ell_{\mathcal{D}}}
  +(\text{lower order terms}),\n
  &\quad c_{\mathcal{D}}^{\Xi}(\bm{\lambda})=
  \prod_{j=1}^{M_{\I}}c_{d^{\I}_j}
  \bigl(\mathfrak{t}^{\I}(\bm{\lambda})\bigr)\cdot
  \prod_{j=1}^{M_{\II}}c_{d^{\II}_j}
  \bigl(\mathfrak{t}^{\II}(\bm{\lambda})\bigr)\cdot
  \prod_{1\leq j<k\leq M_{\I}}(d^{\I}_k-d^{\I}_j)\cdot
  \prod_{1\leq j<k\leq M_{\II}}(d^{\II}_k-d^{\II}_j)\n
  &\quad\phantom{c_{\mathcal{D}}^{\Xi}(\bm{\lambda})=}\times
  \prod_{j=1}^{M_{\I}}\prod_{k=1}^{M_{\II}}
  (a_1+a_1^*-d^{\I}_j-a_2-a_2^*+d^{\II}_k),
  \label{cXiD}\\
  &P_{\mathcal{D}}(\eta;\bm{\lambda})
  =c_{\mathcal{D},n}^{P}(\bm{\lambda})\eta^{\ell_{\mathcal{D}}+n}
  +(\text{lower order terms}),\n
  &\quad c_{\mathcal{D},n}^{P}(\bm{\lambda})=
  c_{\mathcal{D}}^{\Xi}(\bm{\lambda})c_n(\bm{\lambda})
  \prod_{j=1}^{M_{\I}}(-a_1-a_1^*-n+d^{\I}_j+1)\cdot
  \prod_{j=1}^{M_{\II}}(-a_2-a_2^*-n+d^{\II}_j+1).
  \label{cPDn}
\end{align}

\noindent
$\bullet$ $\check{P}_{\mathcal{D},0}(x;\bm{\lambda})$ vs
$\check{\Xi}_{\mathcal{D}}(x;\bm{\lambda})$:
\begin{align}
  &\check{P}_{\mathcal{D},0}(x;\bm{\lambda})
  =A\,\check{\Xi}_{\mathcal{D}}(x;\bm{\lambda}+\bm{\delta}),\n
  &A=\prod_{j=1}^{M_{\I}}(-a_1-a_1^*+d^{\I}_j+1)
  \cdot\prod_{j=1}^{M_{\II}}(-a_2-a_2^*+d^{\II}_j+1).
  \label{PD0=A.XiD}
\end{align}

\noindent
$\bullet$ $d_j=0$ case :
\begin{align}
  &\check{P}_{\mathcal{D},n}(x;\bm{\lambda})\Bigm|_{d^{\I}_{M_{\I}}=0}
  =A\,\check{P}_{\mathcal{D}',n}(x;\bm{\lambda}+\tilde{\bm{\delta}}^{\I}),\n
  &\quad\mathcal{D}'=\{d^{\I}_1-1,\ldots,d^{\I}_{M_{\I}-1}-1,
  d^{\II}_1+1,\ldots,d^{\II}_{M_{\II}}+1\},\n
  &\quad A=(-1)^{M_{\I}}(a_1+a_1^*+n-1)
  \prod_{j=1}^{M_{\I}-1}(-a_1-a_1^*+a_2+a_2^*+d^{\I}_j+1)\cdot
  \prod_{j=1}^{M_{\II}}(d^{\II}_j+1),
  \label{dIM1=0}\\
  &\check{P}_{\mathcal{D},n}(x;\bm{\lambda})\Bigm|_{d^{\II}_{M_{\II}}=0}
  =B\,\check{P}_{\mathcal{D}',n}(x;\bm{\lambda}+\tilde{\bm{\delta}}^{\II}),\n
  &\quad\mathcal{D}'=\{d^{\I}_1+1,\ldots,d^{\I}_{M_{\I}}+1,
  d^{\II}_1-1,\ldots,d^{\II}_{M_{\II}-1}-1\},\n
  &\quad B=(-1)^M(a_2+a_2^*+n-1)
  \prod_{j=1}^{M_{\II}-1}(-a_2-a_2^*+a_1+a_1^*+d^{\II}_j+1)\cdot
  \prod_{j=1}^{M_{\I}}(d^{\I}_j+1).
  \label{dIIM2=0}
\end{align}

%%%%%%%%%%%%%%%%%%%%%%%%%%%%%%%%%%%%%%%%%%%%%%%%%%%%%%%%%%%%%%%
%                                                             %
%  B. Some Properties of the Multi-indexed                    %
%     Meixner-Pollaczek Polynomials                           %
%                                                             %
%%%%%%%%%%%%%%%%%%%%%%%%%%%%%%%%%%%%%%%%%%%%%%%%%%%%%%%%%%%%%%%
\section{Some Properties of the Multi-indexed Meixner-Pollaczek Polynomials}
\label{app:MP:prop}

We present some properties of the multi-indexed Meixner-Pollaczek polynomials.

\noindent
$\bullet$ coefficients of the highest degree terms:
\begin{align}
  &\Xi_{\mathcal{D}}(\eta;\bm{\lambda})
  =c_{\mathcal{D}}^{\Xi}(\bm{\lambda})\eta^{\ell_{\mathcal{D}}}
  +(\text{lower order terms}),\n
  &\quad c_{\mathcal{D}}^{\Xi}(\bm{\lambda})=
  \prod_{j=1}^Mc_{d_j}\bigl(\mathfrak{t}(\bm{\lambda})\bigr)\cdot
  \prod_{1\leq j<k\leq M}(d_k-d_j),
  \label{MP:cXiD}\\
  &P_{\mathcal{D}}(\eta;\bm{\lambda})
  =c_{\mathcal{D},n}^{P}(\bm{\lambda})\eta^{\ell_{\mathcal{D}}+n}
  +(\text{lower order terms}),\n
  &\quad c_{\mathcal{D},n}^{P}(\bm{\lambda})=
  c_{\mathcal{D}}^{\Xi}(\bm{\lambda})c_n(\bm{\lambda})
  \prod_{j=1}^M(-2a-n+d_j+1).
  \label{MP:cPDn}
\end{align}

\noindent
$\bullet$ $\check{P}_{\mathcal{D},0}(x;\bm{\lambda})$ vs
$\check{\Xi}_{\mathcal{D}}(x;\bm{\lambda})$:
\begin{equation}
  \check{P}_{\mathcal{D},0}(x;\bm{\lambda})
  =A\,\check{\Xi}_{\mathcal{D}}(x;\bm{\lambda}+\bm{\delta}),\quad
  A=\prod_{j=1}^M(-2a+d_j+1).
  \label{MP:PD0=A.XiD}
\end{equation}

\noindent
$\bullet$ $d_j=0$ case :
\begin{align}
  &\check{P}_{\mathcal{D},n}(x;\bm{\lambda})\Bigm|_{d_M=0}
  =A\,\check{P}_{\mathcal{D}',n}(x;\bm{\lambda}+\tilde{\bm{\delta}}),\quad
  \mathcal{D}'=\{d_1-1,\ldots,d_{M-1}-1\},\n[4pt]
  &\quad A=(-1)^M(2a+n-1)(2\sin\phi)^{M-1}.
  \label{MP:dM=0}
\end{align}

%%%%%%%%%%%%%%%%%%%%%%%%%%%%%%%%%%%%%%%%%%%%%%%%%%%%%%%%%%%%%%%
%                                                             %
%  References                                                 %
%                                                             %
%%%%%%%%%%%%%%%%%%%%%%%%%%%%%%%%%%%%%%%%%%%%%%%%%%%%%%%%%%%%%%%


\begin{thebibliography}{99}
% 
% for hyphenation : \hspace{0pt}

%%%%%% rdQM 
\bibitem{os12}
S.\,Odake and R.\,Sasaki,
``Orthogonal Polynomials from Hermitian Matrices,''
J. Math. Phys. {\bf 49} (2008) 053503 (43pp),
{\tt arXiv:0712.4106[math.CA]}.
%(The dual $q$-Meixner polynomial in \S\,5.2.4 and dual $q$-Charlier
%polynomial in \S\,5.2.8 should be deleted because the hermiticity of
%the Hamiltonian is lost for these two cases.)

%%%%% idQM
\bibitem{os13}
S.\,Odake and R.\,Sasaki,
``Exactly solvable `discrete' quantum mechanics;
shape invariance, Heisenberg solutions,
annihilation-creation operators and coherent states,''
Prog. Theor. Phys. {\bf 119} (2008) 663-700.
{\tt arXiv:0802.1075[quant-ph]}.

%%%%% dQM
\bibitem{os24}
S.\,Odake and R.\,Sasaki,
``Discrete quantum mechanics,'' (Topical Review)
J. Phys. {\bf A44} (2011) 353001 (47pp),
{\tt arXiv:1104.0473[math-ph]}.

%%%%% orthogonal poly.
\bibitem{ismail}
M.\,E.\,H.\,Ismail,
{\it Classical and Quantum Orthogonal Polynomials in One Variable\/},
vol. 98 of Encyclopedia of mathematics and its applications,
Cambridge Univ. Press, Cambridge (2005).

\bibitem{kls}
R.\,Koekoek, P.\,A.\,Lesky and R.\,F.\,Swarttouw,
{\it Hypergeometric orthogonal polynomials and their $q$-analogues,\/}
Springer-Verlag Berlin-Heidelberg (2010).
%%%%%  -------------------- 

%%%%%%%%%% Xl and miop ---------------
%%%%% Xl poly. in QM (1)
\bibitem{gkm08}
D.\,G\'{o}mez-Ullate, N.\,Kamran and R.\,Milson,
``An extension of Bochner's problem: exceptional invariant subspaces,''
J. Approx. Theory {\bf 162} (2010) 987-1006,
{\tt arXiv:\hspace{0pt}0805.\hspace{0pt}3376[math-ph]};

\bibitem{gkm08_2}
D.\,G\'{o}mez-Ullate, N.\,Kamran and R.\,Milson,
``An extended class of orthogonal polynomials defined by a
Sturm-Liouville problem,''
J. Math. Anal. Appl. {\bf 359} (2009) 352-367,
{\tt arXiv:0807.3939[math-\hspace{0pt}ph]}.

\bibitem{q08}
C.\,Quesne,
``Exceptional orthogonal polynomials, exactly solvable potentials
and supersymmetry,''
J. Phys. {\bf A41} (2008) 392001 (6pp),
{\tt arXiv:0807.4087[quant-ph]}.

\bibitem{os16}
S.\,Odake and R.\,Sasaki,
``Infinitely many shape invariant potentials and new orthogonal polynomials,''
Phys. Lett. {\bf B679} (2009) 414-417,
{\tt arXiv:0906.0142[math-ph]}.

\bibitem{os19}
S.\,Odake and R.\,Sasaki,
``Another set of infinitely many exceptional ($X_{\ell}$) Laguerre
polynomials,''
Phys. Lett. {\bf B684} (2010) 173-176,
{\tt arXiv:0911.3442[math-ph]}.

%%%%% Xl poly. in QM (property)
\bibitem{gkm11} %% X1, word `stable' 
D.\,G\'{o}mez-Ullate, N.\,Kamran and R.\,Milson,
``On orthogonal polynomials spanning a non-standard flag,''
Contemp. Math. {\bf 563} (2011) 51-72,
{\tt arXiv:1101.5584[math-ph]}.

%%%%% miop in QM
\bibitem{gkm11_2} %% miop L M=2
D.\,G\'{o}mez-Ullate, N.\,Kamran and R.\,Milson,
``Two-step Darboux transformations and exceptional Laguerre polynomials,''
J. Math. Anal. Appl. {\bf 387} (2012) 410-418,
{\tt arXiv:\hspace{0pt}1103.5724[math-ph]}.

\bibitem{os25}
S.\,Odake and R.\,Sasaki,
``Exactly solvable quantum mechanics and infinite families of
multi-indexed orthogonal polynomials,''
Phys. Lett. {\bf B702} (2011) 164-170,
{\tt arXiv:\hspace{0pt}1105.0508[math-ph]}.
%(Remark: $\tilde{\bm{\delta}}^{\I}$ and $\tilde{\bm{\delta}}^{\II}$ in this
%reference correspond to $-\tilde{\bm{\delta}}^{\I}$ and
%$-\tilde{\bm{\delta}}^{\II}$ in the present paper, respectively.)

%%%%% Xl in idQM
\bibitem{os17}
S.\,Odake and R.\,Sasaki,
``Infinitely many shape invariant discrete quantum mechanical systems
and new exceptional orthogonal polynomials related to the Wilson and
Askey-Wilson polynomials,''
Phys. Lett. {\bf B682} (2009) 130-136,
{\tt arXiv:0909.3668[math-ph]}.

%%%%% miop in idQM
\bibitem{os27}
S.\,Odake and R.\,Sasaki,
``Multi-indexed Wilson and Askey-Wilson polynomials,''
J. Phys. {\bf A46} (2013) 045204 (22pp),
{\tt arXiv:1207.5584[math-ph]}.

%%%%% Xl in rdQM
\bibitem{os23}
S.\,Odake and R.\,Sasaki,
``Exceptional ($X_{\ell}$) ($q$)-Racah polynomials,''
Prog. Theor. Phys. {\bf 125} (2011) 851-870,
{\tt arXiv:1102.0812[math-ph]}.

%%%%% miop in rdQM
\bibitem{os26}
S.\,Odake and R.\,Sasaki,
``Multi-indexed ($q$-)Racah polynomials,''
J. Phys. {\bf A 45} (2012) 385201 (21pp),
{\tt arXiv:1203.5868[math-ph]}.
%For the type \Romannumeral{2} virtual state vector, see
%{\tt arXiv:1203.5868v1}.)

%%%%% Xl Hermite
\bibitem{ggm13}
D.\,G\'{o}mez-Ullate, Y.\,Grandati and R.\,Milson,
``Rational extensions of the quantum harmonic oscillator and exceptional
Hermite polynomials,''
J. Phys. {\bf A47} (2014) 015203 (27pp),
{\tt arXiv:1306.5143[math-ph]}.
%%%%%  -------------------- 

%%%%% Duran
\bibitem{d13}
A.\,J.\,Dur\'{a}n,
``Exceptional Meixner and Laguerre orthogonal polynomials,''
J. Approx. Theory {\bf 184} (2014) 176-208,
{\tt arXiv:1310.4658[math.CA]}.

\bibitem{os35}
S.\,Odake and R.\,Sasaki,
``Multi-indexed Meixner and Little $q$-Jacobi (Laguerre) Polynomials,''
J. Phys. {\bf A50} (2017) 165204 (23pp),
{\tt arXiv:1610.09854[math.CA]}.
%%%%%%%%%% Xl and miop ---------------

%%%%%% sinusoidal
\bibitem{os7}
S.\,Odake and R.\,Sasaki,
``Unified theory of annihilation-creation operators for solvable
(`discrete') quantum mechanics,''
J. Math. Phys. {\bf 47} (2006) 102102 (33pp),
{\tt arXiv:\hspace{0pt}quant-ph/0605215}.
%
%\bibitem{os8}
%S.\,Odake and R.\,Sasaki,
%``Exact solution in the Heisenberg picture and annihilation-creation
%operators,"
%Phys. Lett. {\bf B641} (2006) 112-117,
%{\tt arXiv:quant-ph/0605221}.
%%%%%  -------------------- 

%%%%% unified (Q)ES
\bibitem{os14}
S.\,Odake and R.\,Sasaki,
``Unified theory of exactly and quasi-exactly solvable `discrete'
quantum mechanics: I. Formalism,"
J. Math. Phys {\bf 51} (2010) 083502 (24pp),
{\tt arXiv:\hspace{0pt}0903.2604[math-ph]}.
%%%%%  -------------------- 

%%%%% crum, adler
\bibitem{crum}
M.\,M.\,Crum,
``Associated Sturm-Liouville systems,"
Quart. J. Math. Oxford Ser. (2) {\bf 6} (1955) 121-127,
{\tt arXiv:physics/9908019}.

\bibitem{krein}
M.\,G.\,Krein,
``On continuous analogue of a formula of Christoffel from the theory
of orthogonal polynomials," (Russian)
Doklady Acad. Nauk. CCCP, {\bf 113} (1957) 970-973.

\bibitem{adler}
V.\,\'E.\,Adler,
``A modification of Crum's method,''
Theor. Math. Phys. {\bf 101} (1994) 1381-1386.
%%%%%  -------------------- 

%%%%% dQM
\bibitem{os15}
S.\,Odake and R.\,Sasaki,
``Crum's theorem for `discrete' quantum mechanics,''
Prog. Theor. Phys. {\bf 122} (2009) 1067-1079,
{\tt arXiv:0902.2593[math-ph]}.

\bibitem{gos}
L.\,Garc\'ia-Guti\'errez, S.\,Odake and R.\,Sasaki,
``Modification of Crum's theorem for `discrete' quantum mechanics,''
Prog. Theor. Phys. {\bf 124} (2010) 1-26,
{\tt arXiv:1004.0289\hspace{0pt}[math-ph]}.
%%%%%  -------------------- 

%%%%% typeIII, other potentials
\bibitem{q12b}  %% RM, Eckart
C.\,Quesne,
``Novel enlarged shape invariance property and exactly solvable rational
extensions of the Rosen-Morse II and Eckart potentials,"
SIGMA {\bf 8} (2012) 080 (19pp),
{\tt arXiv:1208.6165[math-ph]}.

%%%%% typeIII, other potentials
\bibitem{os29}
S.\,Odake and R.\,Sasaki,
``Krein-Adler transformations for shape-invariant potentials and pseudo
virtual states,"
J. Phys. {\bf A46} (2013) 245201 (24pp),
{\tt arXiv:1212.6595[math-\hspace{0pt}ph]}.

\bibitem{os28}
S.\,Odake and R.\,Sasaki,
``Extensions of solvable potentials with finitely many discrete eigenstates,"
J. Phys. {\bf A46} (2013) 235205 (15pp),
{\tt arXiv:1301.3980[math-ph]}.

%%%%% Xl, miop in idQM
\bibitem{os30}
S.\,Odake and R.\,Sasaki,
``Casoratian Identities for the Wilson and Askey-Wilson Polynomials,"
J. Approx. Theory {\bf 193} (2015) 184-209,
{\tt arXiv:1308.4240[math-ph]}.

%%%%% miop in rdQM
\bibitem{os22}
S.\,Odake and R.\,Sasaki,
``Dual Christoffel transformations,''
Prog. Theor. Phys. {\bf 126} (2011) 1-34,
{\tt arXiv:1101.5468[math-ph]}.

%%%%% equivalence
\bibitem{equiv_miop}
S.\,Odake,
``Equivalences of the Multi-Indexed Orthogonal Polynomials,"
J. Math. Phys. {\bf 55} (2014) 013502 (17pp),
{\tt arXiv:1309.2346[math-ph]}.

\bibitem{t13}
K.\,Takemura,
``Multi-indexed Jacobi polynomials and Maya diagrams,''
J. Math. Phys. {\bf 55} (2014) 113501 (10pp), 
{\tt arXiv:1311.3570[math-ph]}.
%%%%%  --------------------

%%%%% Xl poly. in QM (property)
\bibitem{stz10}
R.\,Sasaki, S.\,Tsujimoto and A.\,Zhedanov,
``Exceptional Laguerre and Jacobi polynomials and the corresponding
potentials through Darboux-Crum transformations,''
J. Phys. {\bf A43} (2010) 315204,
{\tt arXiv:1004.4711[math-ph]}.

%%%%% 3+2M term
\bibitem{rrmiop}
S.\,Odake,
``Recurrence Relations of the Multi-Indexed Orthogonal Polynomials,''
J. Math. Phys. {\bf 54} (2013) 083506 (18pp),
{\tt arXiv:1303.5820[math-ph]}.

%%%%% duran
\bibitem{d14_2}
A.\,J.\,Dur\'{a}n,
``Higher order recurrence relation for exceptional Charlier, Meixner,
Hermite and Laguerre orthogonal polynomials,''
Integral Transforms Spec. Funct. {\bf 26} (2015) 357-376,
{\tt arXiv:1409.4697[math.CA]}.

%%%%% 3+2l term
\bibitem{mt14}
H.\,Miki and S.\,Tsujimoto,
``A new recurrence formula for generic exceptional orthogonal polynomials,''
J. Math. Phys. {\bf 56} (2015) 033502 (13pp),
{\tt arXiv:1410.0183[math.CA]}.

\bibitem{rrmiop2}
S.\,Odake,
``Recurrence Relations of the Multi-Indexed Orthogonal Polynomials : $\II$,''
J. Math. Phys. {\bf 56} (2015) 053506 (18pp),
{\tt arXiv:1410.8236[math-ph]}.
%(typo in eq.(B.12): RHS of $X(\eta)$,
%$\frac{\eta}{(1+q)\sigma_1}\rightarrow\frac{\eta}{(1+q)\sigma_2}$)
%
%%%%%% web page
%\bibitem{wp}
%\verb|http:|\verb|//azusa.shinshu-u.ac.jp/~odake/paper/data/1410.8236.html|.
%%%%%%  --------------------

%%%%%% rr Hermite
\bibitem{gkkm15}
D.\,G\'{o}mez-Ullate, A.\,Kasman, A.\,B.\,J.\,Kuijlaars and R.\,Milson,
``Recurrence Relations for Exceptional Hermite Polynomials,''
J. Approx. Theory {\bf 204} (2016) 1-16,
{\tt arXiv:\hspace{0pt}1506.03651[math.CA]}.
%%%%%  --------------------

%%%%% 3+2l term proof
\bibitem{rrmiop3}
S.\,Odake,
``Recurrence Relations of the Multi-Indexed Orthogonal Polynomials : $\III$,''
J. Math. Phys. {\bf 57} (2016) 023514 (24pp),
{\tt arXiv:1509.08213[math-ph]}.
%%%%%  --------------------

%%%%% 3+2l term and closure relation
\bibitem{rrmiop4}
S.\,Odake,
``Recurrence Relations of the Multi-Indexed Orthogonal Polynomials $\IV$ :
closure relations and creation/annihilation operators,''
J. Math. Phys. {\bf 57} (2016) 113503 (22pp),
{\tt arXiv:1606.0283[math-ph]}.
%%%%%  --------------------

%%%%% rr for qR
\bibitem{rrmiop5}
S.\,Odake,
``Recurrence Relations of the Multi-Indexed Orthogonal Polynomials $\V$ :
Racah and $q$-Racah types,''
J. Math. Phys. {\bf 60} (2019) 023508 (30pp),
{\tt arXiv:1804.10352\hspace{0mm}[math-ph]}.
%%%%%  --------------------

%%%%% rr for cH,MP
\bibitem{rrmiop6}
S.\,Odake,
``Recurrence Relations of the Multi-Indexed Orthogonal Polynomials $\VI$ :
Meixner-Pollaczek and continuous Hahn types,''
in preparation.
%%%%%  --------------------

\end{thebibliography}
\end{document}